\documentclass[times,authoryear]{elsarticle}

\usepackage{jasr}
\usepackage{framed,multirow}

\usepackage{amssymb}
\usepackage{latexsym}

\usepackage[switch]{lineno}

\usepackage{url}
\usepackage{xcolor}
\definecolor{newcolor}{rgb}{.8,.349,.1}

\usepackage[citebordercolor=white]{hyperref}

\journal{ASR}

\begin{document}

\verso{Chakraborty and Seemala}

\begin{frontmatter}

\title{Ionospheric responses over the Antarctic region to Intense Space Weather events: Plasma Convection vs. Auroral Precipitation}

\author[a]{Sumanjit Chakraborty\corref{c-d54cc1eb1ca4}}
\ead{sumanjit11@gmail.com}\cortext[c-d54cc1eb1ca4]{Corresponding author.}
\author[a]{Gopi K. Seemala}
\ead{gopi.seemala@gmail.com}

\affiliation[1]{organization={Indian Institute of Geomagnetism}, 
                postcode={Navi Mumbai 410218},
                country={Maharashtra, India}
                }

\received{xx}
\finalform{xx}
\accepted{xx}
\availableonline{xx}
\communicated{}

\begin{abstract}

The present investigation is directed at exploring southern polar ionospheric responses to intense space weather events and their correlations with plasma convection and auroral precipitation. The main phases of six geomagnetic storms occurring in the year 2023 (ascending phase of the present solar cycle) are considered for this study. The ionospheric Total Electron Content (TEC) measurements derived from GPS receivers covering the Antarctic region are used for probing the electron density perturbations during these events. Auroral precipitation maps are shown to illustrate the locations of the GPS stations relative to particle precipitation. SuperDARN maps are shown to understand the effects of plasma convection over these locations. Correlation between the enhanced TEC observations with the auroral precipitation (R $\sim$ 0.31) and the plasma convection (R $\sim$ 0.88) reveals that the latter is more responsible for causing significant enhancements in the diurnal maximum values of TEC over the Antarctic region in comparison to the former. Therefore, this work shows correlation studies between two physical processes and ionospheric density enhancements over the under-explored south polar region under strong levels of geomagnetic activity during 2023.  

\end{abstract}

\begin{keyword}
\KWD Ionospheric TEC  \sep Geomagnetic Storms \sep Plasma Convection \sep Antarctic Stations \sep GPS \sep Auroral Precipitation
\end{keyword}

\end{frontmatter}


\section{Introduction}

Geomagnetic storms, manifested as large disturbances of the Earth's magnetic field, occur mainly following Coronal Mass Ejections (CMEs) from the Sun. When the north-south ($B_z$) component of the frozen-in Interplanetary Magnetic Field (IMF) becomes fully southward, geomagnetic storm-time conditions prevail \citep{sc:01,sc:02}. When the $B_z$ stays in this orientation for a sufficient time interval (at least 3 hours), strong/intense (SYM-H $\leq$ -100 nT) geomagnetic storms occur \citep{sc:03,sc:04,sc:05,sc:06}. The polar region is highly sensitive to these geomagnetic storms \citep{sc:22}, coupled with the strongest and most frequent occurrence of ionospheric irregularities and associated scintillations affecting satellite-based navigation and communication systems. This makes polar research an important part of modern-day society that is highly dependent on the technological systems \citep{sc:17,sc:18,sc:19,sc:20,sc:21}. 

An important parameter to gauge the ionosphere is the Total Electron Content (TEC). It is the electron density integrated along the radio signal path from the satellite to the receiver and is expressed in TEC Units or TECU (1 TECU = 10$^{16}$ electrons/$m^2$). The TEC varies strongly with varying levels of geomagnetic activity (see \cite{sc:23,sc:24,sc:25,sc:26,sc:27,sc:28} and references therein). The understanding and investigation of changes and subsequent modeling of the ionospheric TEC becomes crucial when one uses TEC for the forecasting of ionospheric delays and associated navigational errors for the improvement of GNSS/GPS performance, especially under strong geomagnetic storm-time conditions.  

The ionosphere over the southern polar region has attracted scientific interest due to its geophysical characteristics and the influence of high-latitude physical processes on the ionosphere, different from its northern counterpart. The TEC is known to exhibit geomagnetic variations (in addition to diurnal and seasonal variabilities) driven by the interaction between the solar wind, magnetospheric convection, and thermospheric dynamics. Ground-based observations from GPS, as well as in-situ observations from satellites such as DMSP, COSMIC/FormoSat-3, etc, have provided useful insights into the Antarctic ionospheric variabilities. \cite{sc:a} demonstrated the possible role of polar cap convection, specifically subauroral polarization streams, in modulating the TEC during geomagnetic storm-time conditions. In the study by \citep{sc:f}, TEC enhancements have been shown over the southern hemisphere. They stated that under southward IMF $B_z$, TEC fluctuations increase in comparison to when under northward IMF conditions. The study by \citep{sc:g} showed variability in the ionosphere over the Indian Antarctic station Bharati and stated that enhancement/suppression of TEC under the influence of geomagnetic storms depends on the storm onset time. In a recent study, \citep{sc:b} identified the polar cap patches or regions of enhanced ionization appearing under southward IMF \citep{sc:c} using ground-based TEC observations and in-situ ionospheric parameters from the DMSP. Furthermore, initial studies by \citep{sc:d,sc:e} show observations related to plasma convection patterns and auroral precipitations over the Antarctic region. 

Despite these studies, the Antarctic region still remains less explored due to the sparse or limited number of ground-based receiver stations over the entire region \citep{sc:29}, unlike its Arctic counterpart, where there exist several ground-based radars and experiments having good spatial and temporal coverage. Several studies (see \cite{sc:30,sc:31,sc:16,sc:32} and references therein) focus on describing the northern polar ionosphere under both disturbed and quiet-time conditions. Additionally, due to the presence of a large offset between the geomagnetic and the geographic poles, the changes in the Solar wind-magnetosphere-ionosphere (SwMI) coupling would cause different responses in the southern polar regions in comparison to their northern counterparts, even if the driving mechanisms are the same (see \cite{sc:41,sc:42} and references therein). Hence, it becomes interesting to observe the southern polar region dynamics under different space weather events.

Given the requirement to study the polar ionosphere under geomagnetically active conditions, focusing on the under-explored Antarctic region, the present investigation shows the ionospheric response in terms of variations in the TEC during six intense geomagnetic storms (see Table \ref{tab1}) over the entire year of 2023 (ascending-to-maximum phase of the present solar cycle 25). This study aims to explore the ionospheric responses over Antarctica in terms of the correlation of the plasma convection velocities and auroral precipitation densities, to the enhanced TEC variations during the main phases of strong geomagnetic storm-time conditions that had occurred in the year 2023.

The manuscript is divided as follows: Section 2 describes the data used and the corresponding methodology. Section 3 presents detailed results of a storm as a case study from the storm database selected, as well as statistical variation for all the events. Section 4 presents the discussion, while Section 5 summarizes the present investigation. 

\begin{table*}
\centering
\vspace{20pt}
\caption{Strong geomagnetic storm events of 2023. It is to be noted that the first three cases fall around the equinox, while the latter three are around the solstice.}
\vspace{10pt}
\begin{tabular}{|l|l|l|l|}
\hline
Event number & Event period & minimum SYM-H (nT) & UT (h), Date (DD)\\
\hline
GS1 & February 26-28  & -161 & 12:12, 27 \\
\hline
GS2 & March 23-25     & -170 & 05:21, 24 \\
\hline
GS3 & April 23-25     & -233 & 04:03, 24 \\
\hline
GS4 & November 04-06  & -189 & 16:54, 05 \\
\hline
GS5 & November 24-26  & -109 & 19:05, 25 \\
\hline
GS6 & December 01-03  & -136 & 13:30, 01 \\
\hline
\end{tabular}
\label{tab1}
\end{table*}

\section{Data and Methodology}

Since 1989 and 2013, two Antarctic research stations, Maitri (mtri) and Bharati (bhrt), respectively, have been operational. They are handled by the National Centre for Polar and Ocean Research (NCPOR), Ministry of Earth Sciences (MoES), Government of India. The TEC measurements from the GPS receivers at these two stations, along with other available stations over the entire Antarctic region, are considered for this study. The GPS station list with the corresponding geographic and geomagnetic coordinates is presented in Table \ref{tab2}.

\begin{table*}
\centering
\vspace{20pt}
\caption{List of GPS stations (arranged alphabetically) and the corresponding geographic and geomagnetic coordinates. over the Antarctic region}
\vspace{10pt}
\begin{tabular}{|l|l|l|}
\hline
Station & Geographic Lat, Long & Geomagnetic Lat, Long\\
\hline
Bharati     (bhrt)   & (-69.41, 76.19)  & (-76.38, 129.01) \\
\hline
Maitri      (mtri)   & (-70.77, 11.73)  & (-67.82, 60.86)  \\ 
\hline
Palmer      (palm)   & (-64.78, -64.05) & (-55.44, 6.49)   \\
\hline
Scott Base  (sctb)   & (-77.85, 166.76) & (-79.09, -74.40) \\
\hline
\end{tabular}
\label{tab2}
\end{table*}

For the two Indian Antarctic stations, GPS receivers (Leica 1200) were used to collect phase data and raw pseudorange from the GPS observables over the two Indian stations. It was then converted to Receiver INdependent EXchange (RINEX) format using the TEQC program \citep{sc:61}. Next, we used the GPS TEC program \citep{sc:43,sc:64} that read these RINEX observation data files obtained from the stations and the International GNSS Service (IGS) network of ground-based GPS receivers. The final Vertical TEC (VTEC) was then calculated using this GPS TEC software \citep{sc:43,sc:64} where a single shell (assuming the Ionospheric Pierce Point at 350 km altitude) mapping function \citep{sc:44,sc:45,sc:46} gets used for the conversion of Slant TEC (STEC) to VTEC. The complete methodology (including removal of satellites and receiver biases before obtaining the final VTEC from STEC) is detailed in \citep{sc:64} and under Section 2 of \citep{sc:29}.

Additionally, openly available 1-minute, high-resolution IMF components $B_z$ and $B_y$, the solar wind velocity ($V_{sw}$), and the solar wind density, the Interplanetary Electric Field ($IEF_y$), along with the SYM-H variation, are obtained from the OMNIWeb database of SPDF, GSFC, NASA. The westward auroral electrojet (SML) data is obtained from the SuperMag network. The SuperMag is a global collaboration of national organizations and agencies operating more than 300 ground-based magnetometers. They provide magnetic field variations in a common coordinate system, with the same time resolution and a common baseline removal approach \citep{sc:53,sc:54}.

Moreover, we have used TEC variations from the NCAR Whole Atmosphere Community Climate Model with thermosphere and ionosphere eXtension (WACCM-X) simulations. It is a general circulation model that fully couples the atmospheric chemistry and the dynamics and is self-consistent. It calculates three-dimensional temperature, composition, wind, and ionospheric structures from the surface up to an altitude of 700 km. The model inputs are high-latitude ionospheric inputs and solar spectral irradiance. The outputs of this model are the temperatures and densities of electrons and ions, the Hall and Pederson conductivities, the meridional, the zonal, and the vertical ion drifts and neutral winds in addition to compositions such as $H$, $NO$, $O$, $O_2$, etc (see \cite{sc:47,sc:48} and references therein). 

To understand the nature of auroral precipitation over the polar region, we have used the Ovation-Prime model \citep{sc:59,sc:60} developed by Patrick Newell of the Johns Hopkins University Applied Physics Laboratory (JHUAPL). This model was developed using energetic particle measurements from the Defense Meteorological Satellite Program (DMSP) satellites. It provides the statistical distribution of auroral precipitation obtained from 11 years of electrostatic analyzer data from the DMSP. 

Finally, we have used the ionospheric convection maps from one of the Super Dual Auroral Radar Network (SuperDARN) statistical convection models. We used the TS18 \citep{sc:62} model where inputs are the IMF $B_y$ and $B_z$, the $V_{sw}$, and the date and time. The SuperDARN consists of over 30 HF radars of low power. They are capable of observing the plasma dynamics in the upper atmosphere over the polar regions to the mid-latitudes \citep{sc:56,sc:57,sc:58}.   

\section{Results}

In this section, we show the ionospheric variations over the southern polar region under the strong geomagnetic storm-time conditions during March 23-25, 2023, in detail, followed by the statistical observations of all six events under consideration. 

According to the National Oceanic and Atmospheric Administration-Space Weather Prediction Center (NOAA-SWPC), a G3 level or strong geomagnetic storm was observed on March 23, 2023, at 14:49 UT as a result of a CME, ejected from the Sun on March 20, 2023. Further, due to the presence of the effects of CH-HSSWs and a CIR, geomagnetic storm (G1-G2: minor-to-moderate) effects persisted from March 24 through March 25, 2023. The effects of such a combination of CIRs and CH-HSSWs are known to cause drastic changes to the upper atmosphere over the polar regions, the mid-latitudes, as well as the low-to-equatorial latitudes (see \cite{sc:49} and references therein)      

\begin{figure*}
\centering
\includegraphics[scale=0.81]{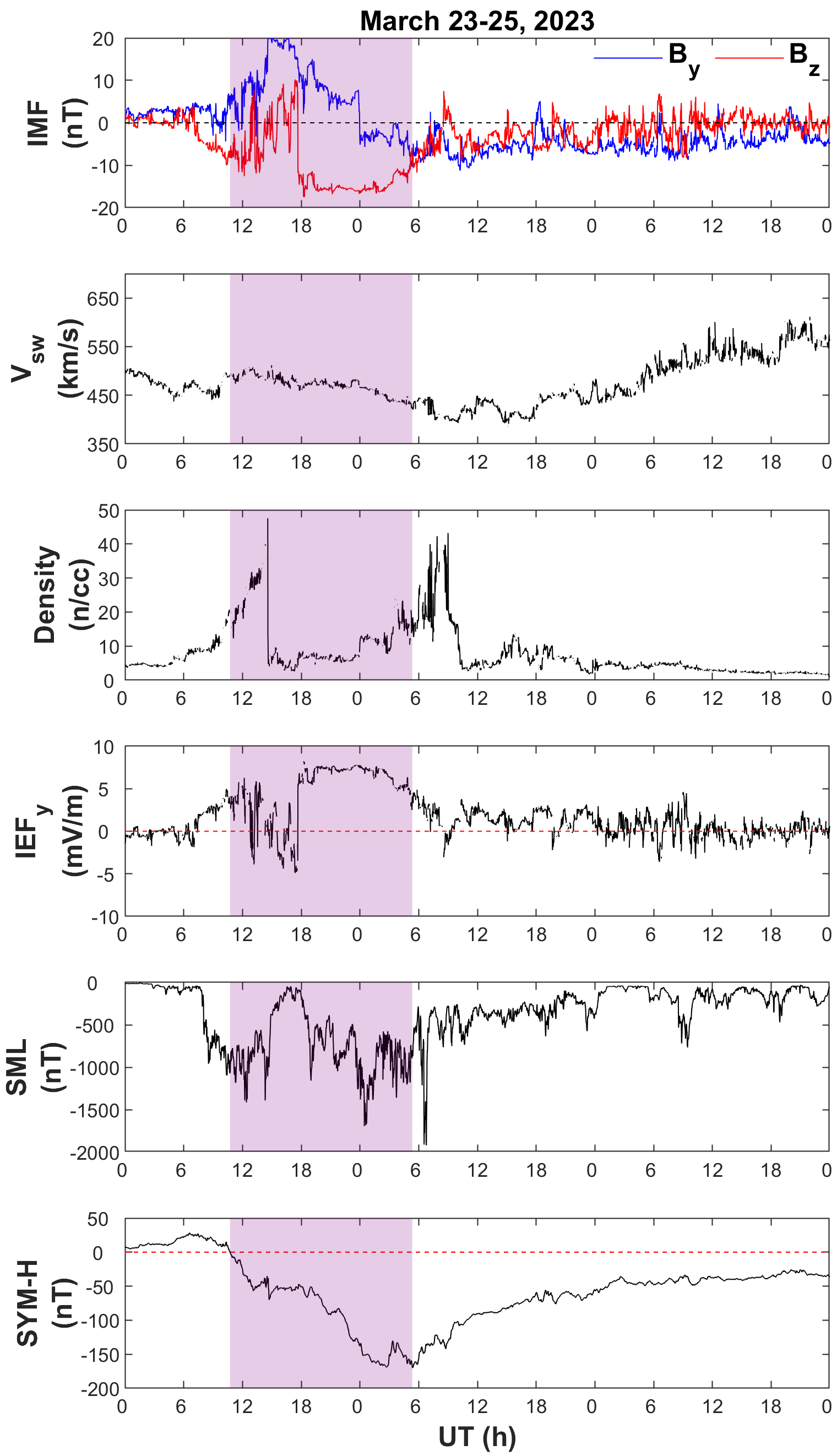}
\caption{From Top to bottom: variations in the IMF (nT) $B_z$ (red) and $B_y$ (blue), the $V_{sw}$ (km/s), the density (n/cc), the $IEF_y$ (mV/m) along with the westward auroral electrojet SML and the SYM-H variations (nT) during March 23-25, 2023. The peach-shaded region shows the main phase of this storm.}
\label{fig01}
\end{figure*}

Figure \ref{fig01} shows the solar wind and the IMF conditions, along with the magnetic field perturbations over the ground. The peach-shaded region signifies the Main Phase (MP) of the geomagnetic storm. It is to be noted that the MPs are selected as per the supporting information (Data Set S2) of \citep{sc:50}. The MP onset, as observed from the bottom panel (SYM-H variations), was at 10:44 UT on March 23. The event was negatively double-peaked, having one dip at 02:40 UT and the other dip (marking the end of the MP) at 05:21 UT on March 24, with values of -169 nT and -170 nT, respectively. This event also had an extended recovery phase that ended at 03:57 UT (not shown) on March 26, 2023. At the MP onset time, the $B_z$ (top panel) had been southward with the value of -8.32 nT, while the $B_y$ (the same top panel) had been duskward with the value of 3.90 nT. Next, the $B_z$ started turning northward and was fully northward at 17:22 UT with a peak value of 10.20 nT. It then finally started turning southward and remained completely southward for about 14 hours from 17:39 UT on March 23 to 07:11 UT on March 24, 2023, when it became northward for a minute and turned southward till becoming northward again at 08:31 UT During this timeline, the $B_y$ smoothly turned from duskward to dawnward. The nature of $B_z$ variation showed the presence of a sheath region followed by a magnetic cloud region (see \cite{sc:51} and references therein). Furthermore, the presence of fluctuating $B_z$ \citep{sc:52} during the recovery phase past the shaded region in this figure) confirmed the presence of HSSWs/CIR. Coming to the second panel from the top, one can observe a steady decrease in the solar wind velocity. During the entire MP, the $V_{sw}$ decreased from about 520 km/s to 440 km/s. However, the same can be observed to increase to higher values (about 600 km/s) from the beginning of March 25, 2023, due to the presence of CIR and HSSWs. Coming to the third panel from the top, at the MP onset, the density shows an increase from around 20 n/cc to a first peak of about 48 n/cc and drops back down to around 10 n/cc before rising back to about 17 n/cc at the end of MP. The second peak of this density can be observed to be around 40 n/cc, although in the recovery phase of the storm. As $IEF_y$ (third panel from the bottom) is the cross product of IMF $B_z$ and the $V_{sw}$, one can observe a trend reversal with respect to the $B_z$ variation in the top panel. When observing the ground-based magnetometer measurement (second to bottom panel) of the westward auroral electrojet from the SuperMag, SML shows strong and multiple (values going down to about -1750 nT) substorm activities during the entire MP. The strongest substorm can be observed to be -1921 nT, however, towards the beginning of the recovery phase. Overall, this particular storm had been very dynamic due to the presence of different substructures of CME (sheath, magnetic cloud), followed by CIR and HSSWs.

\begin{figure*}
\centering
\includegraphics[width=\linewidth,height=180pt]{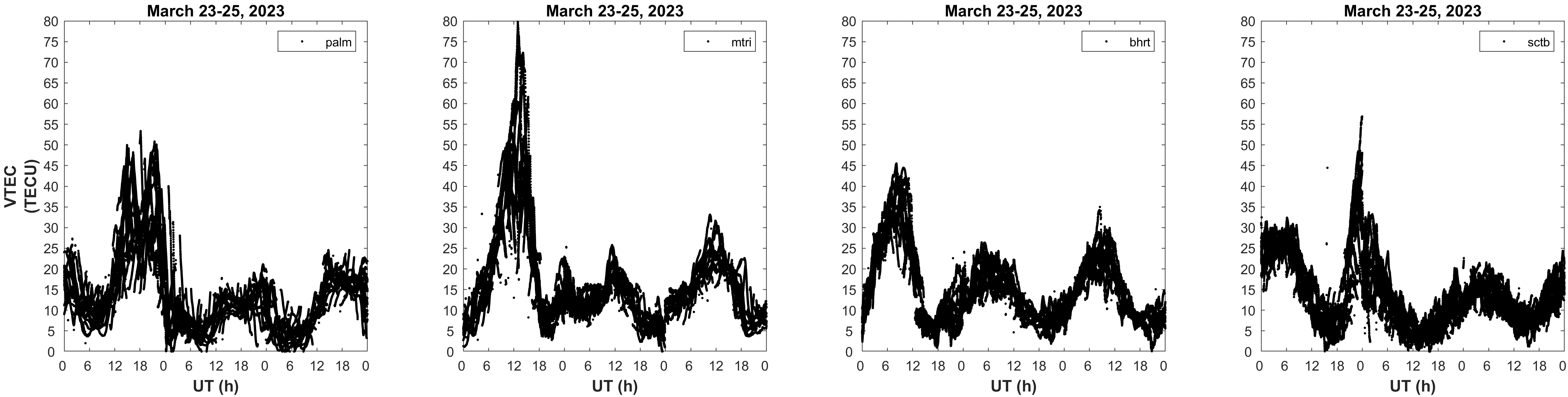}
\includegraphics[width=\linewidth,height=180pt]{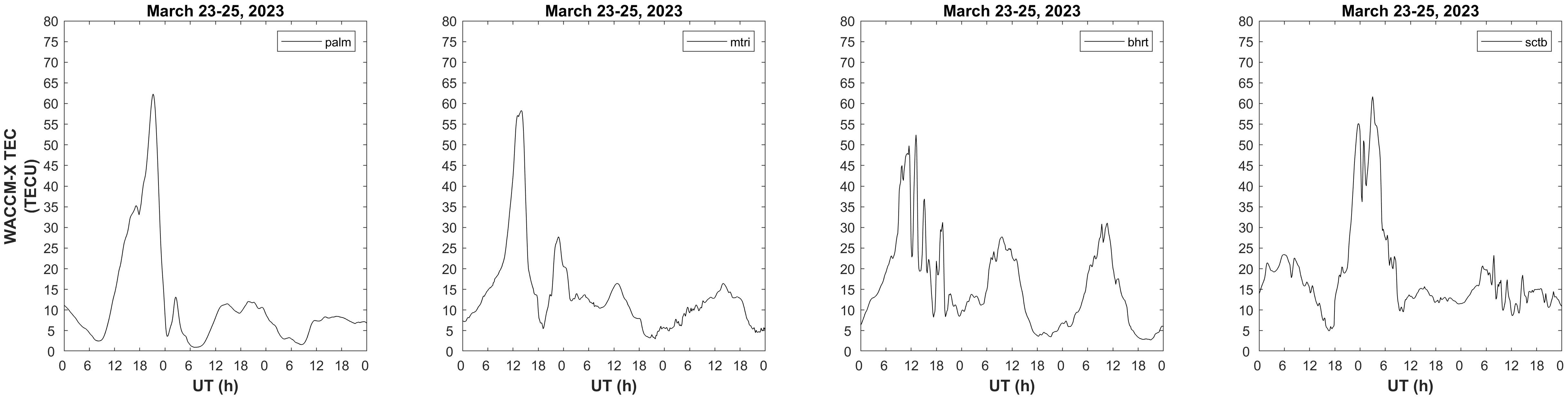}
\caption{Diurnal TEC variations (top panel: GPS-derived observations and bottom panel: WACCM-X simulations) over the four Antarctic stations: palm, mtri, bhrt, and sctb during March 23-25, 2023.}
\label{fig02}
\end{figure*}

Figure \ref{fig02} (top panel) shows the observed GPS-derived VTEC variations over the four Antarctic stations: palm, mtri, bhrt, and sctb, arranged from geographic west to east as one goes from the left panel to the right. The corresponding simulated TEC variations from the WACCM-X model are shown in the bottom panel. Looking at the top panel of this figure, we can observe a highly enhanced diurnal maximum of TEC value of 79.93 TECU at 13 UT on March 23, 2023, over mtri. There are slight enhancements over the stations: palm (at 18 UT on March 23), bhrt (07 UT on March 23), and sctb (00 UT on March 24); however, they are not as high as that observed over mtri. It is to be noted that even though the latitudinal separation between mtri and bhrt is about 1.3$^\circ$, such a level of difference in the diurnal TEC maximum can be observed over the stations. Looking into the bottom panel, the enhancement level of the WACCM-X TEC is almost similar over all four stations, signifying the fact that this model can reproduce TEC enhancements that are expected during a storm MP. However, it is not able to capture the anomalously enhanced TEC variations over mtri.   

\begin{figure*}
\centering
\includegraphics[scale=0.81]{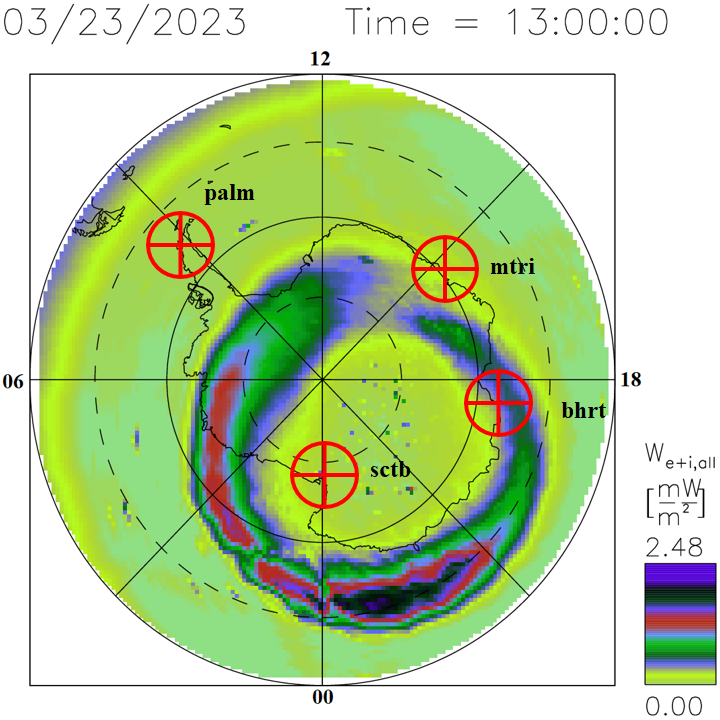}
\caption{Auroral precipitation pattern over the south polar region on March 23, 2023, at 13 UT. The four stations are marked on the map to observe their locations with respect to the auroral oval and particle precipitation.}
\label{fig03}
\end{figure*}

Next, we show Figure \ref{fig03} to understand the nature of the particle (electrons and ions together) precipitation and the location of the auroral oval during the period (13 UT on March 23, 2023) of enhanced TEC variations over mtri. The locations of the four Antarctic stations are marked on the map. It can be observed that the location of the station: mtri is outside the auroral oval and is far away from the peak distribution (about 2.5 $mW/m^2$); however, the same showed TEC enhancements up to 80 TECU. Therefore, the observed enhancement could not be attributed to excess auroral precipitation at this time. 

\begin{figure*}
\centering
\includegraphics[scale=0.81]{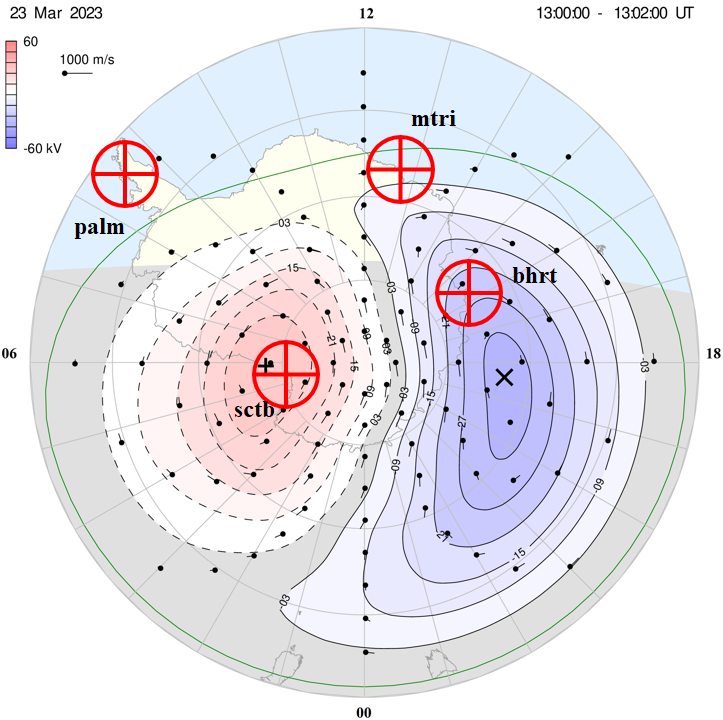}
\caption{Two-cell convection pattern over the south polar region on March 23, 2023, during the 13:00-13:02 UT window. The four stations are marked on the map so that their locations can be observed with respect to the convection pattern. The dawn (red) cell shows positive potential (+ sign), while the dusk (blue) cell shows negative potential (X sign).}
\label{fig04}
\end{figure*}

We further look into the SuperDARN ionospheric convection map (Figure \ref{fig04}) to understand the roles played by plasma convection during this period. The inputs given for this particular model run were the $B_y$ and $B_z$ values, the $V_{sw}$, and the date and time (March 23, 2023, 13 UT). In the context of the two-cell convection pattern's evolution, it is known that under the southward-oriented $B_z$, the dawn (red) cell takes the circular or an orange-shaped structure, while the dusk (blue) cell takes the form of a banana or becomes a crescent-like structure. In between these cells, around the region known as the electrodynamical divider (see \cite{sc:55} and references therein), there exists a distinct anti-sunward throat flows along the noon-midnight meridian. Upon observation of the position of mtri with respect to the convection pattern, we can observe it to be in the vicinity of the anti-sunward throat flows region, which is not the case for the other stations. The station palm is well outside the two-cell convection, while the stations bhrt and sctb are far away from the electrodynamical divider. Therefore, the TEC enhancement observed over mtri can be attributed to the station being around the throat flows region. In the following paragraph, using a similar approach as shown in detail for this event, we show the values of enhancements in the diurnal maximum of TEC for the other events (i.e., GS1 and GS3 to GS6 in Table \ref{tab1}). 

For GS1, the station bhrt showed a diurnal maximum TEC value of 68.78 TECU at 11:30 UT on February 27. For GS3, the same was observed over the station palm with the value of 63.85 TECU at 19:28 UT on April 23. For GS4, the diurnal maxima of TEC peaked at the value of 50.52 TECU over the palm at 18:30 UT on November 05. For GS5 and GS6, the values peaked at 38.12 TECU and 42.42 TECU over mtri at 11:12 UT on November 25 and 12:46 UT on December 01, respectively.

\begin{figure*}
\centering
\includegraphics[scale=0.81]{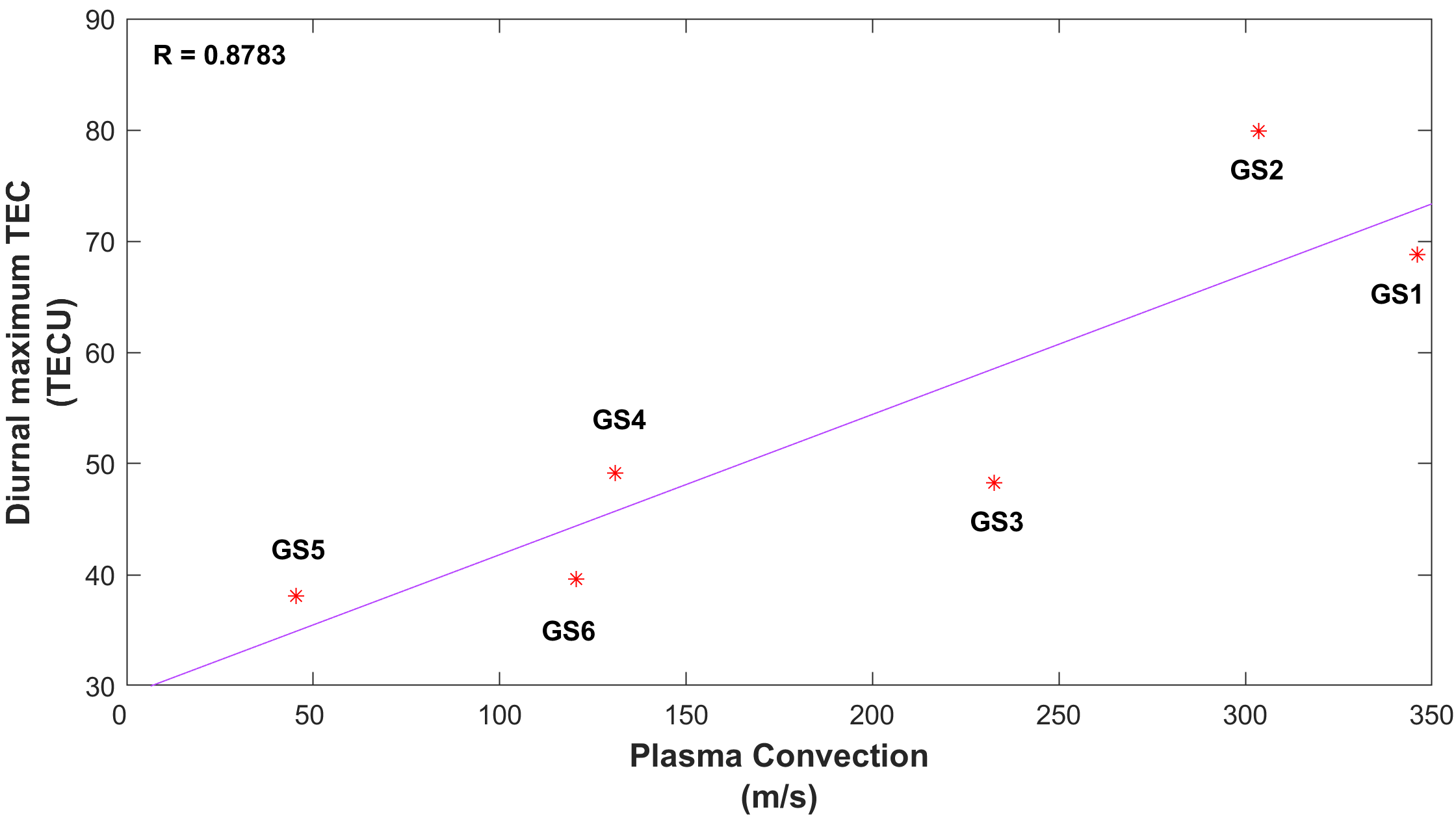}
\includegraphics[scale=0.81]{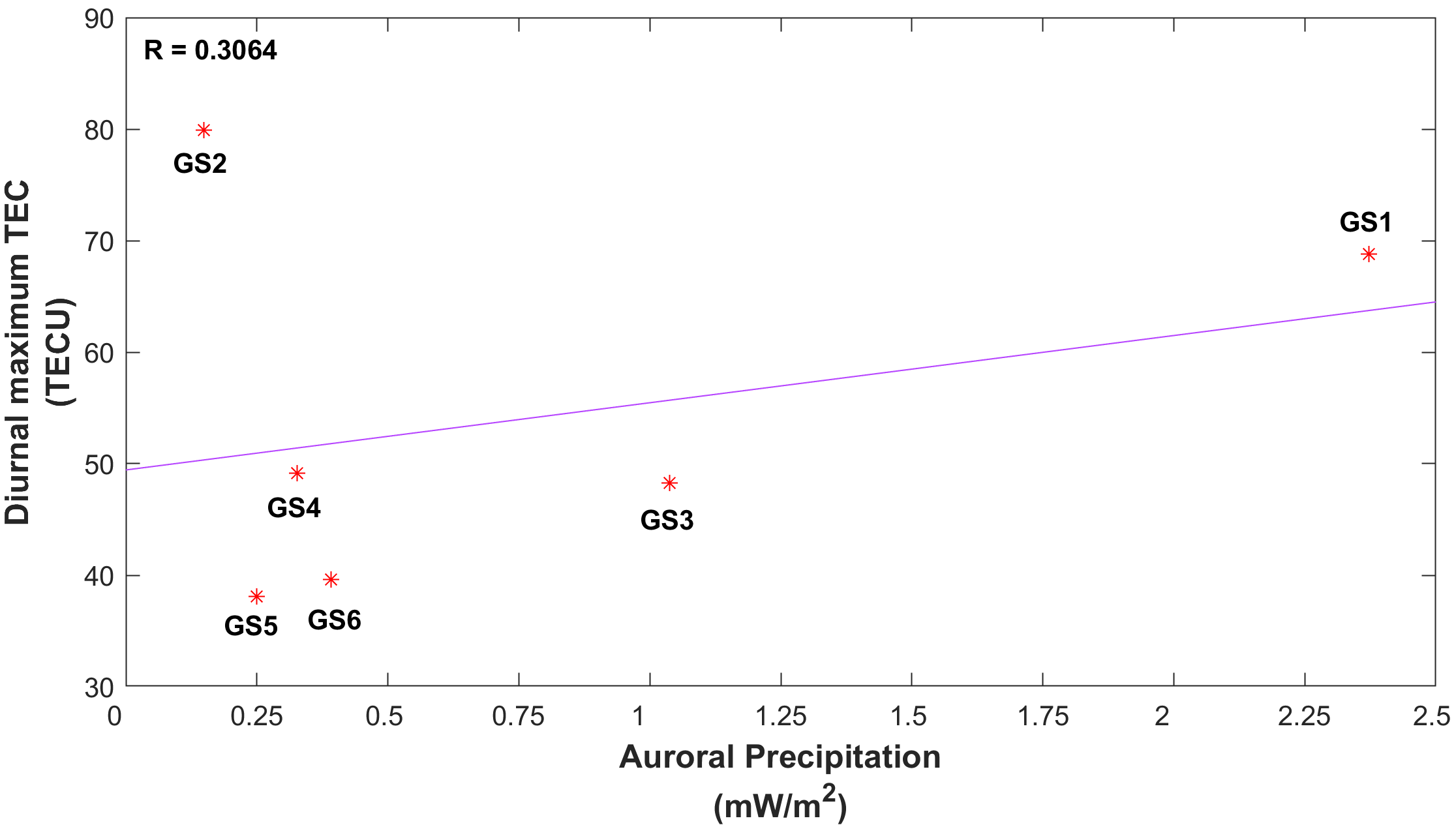}
\caption{Correlation between the plasma convection (top) and the particle precipitation (bottom) with the diurnal maximum of TEC. The six geomagnetic storm (GS1 to GS6) events are designated following the nomenclature adapted in Table \ref{tab1}.}
\label{fig05}
\end{figure*}

Finally, Figure \ref{fig05} shows how the auroral precipitation and the plasma convection are correlated with the observed enhancements in the TEC for all the events (GS1 to GS6). The top panel shows the correlation of the diurnal maximum of TEC with the plasma convection, while the bottom panel shows the same with auroral precipitation. From these two observations, we can see that there is a very good correlation between the observed diurnal maximum of TEC and the plasma convection velocity (R = 0.8783), while there exists no such correlation between the observed TEC variations and the particle precipitation (R = 0.3064). Therefore, from this statistical investigation, it is evident that during the main phase of the six strong geomagnetic storms and in comparison to precipitation, the convection is highly correlated with the observed diurnal maxima of TEC.

\section{Discussion}

The electron density's morphology over the polar and high latitudes differs as a result of asymmetry in the interaction between the magnetosphere and solar wind over the two poles. Hence, one would expect the response of the southern polar ionosphere under storm-time conditions to be different from its northern counterpart. Further, the physical processes considered in the present work exhibit altitudinal as well as spatial domains of dominance within the coupled magnetosphere-ionosphere system. On one hand, precipitation serves as a localized source of ionization, which is mainly confined to auroral latitudes, and its effects are concentrated in the ionospheric regions (D and E layers) of lower altitudes \citep{sc:h,sc:i}. On the other hand, the $E \times B$ convection governs the large-scale plasma transport, thus playing a dominant role in plasma redistribution across global scales. This primarily affects the higher altitude (F-region) ionosphere and plasmasphere \citep{sc:j,sc:k}. Plasma convection at auroral and polar latitudes is driven by magnetospheric processes, which in turn are driven by the solar wind and the IMF conditions that result in a highly variable pattern of convection \citep{sc:01,sc:l,sc:m}. 

During geomagnetic storm-time conditions and subsequent southward configuration of the north-south ($B_z$) component of the IMF, the dawn and the dusk cells respectively take the forms of an orange (nearly circular) and a banana (crescent-like), separated by the electrodynamical divider and with the specialty of the presence of distinct anti-sunward throat flows in between the two distorted cells. During the geomagnetic storm's main phase, the cross-polar convection's two-cell configuration is determined by the southward turned $B_z$. However, the shape of these two cells is mainly determined by the polarity of the IMF $B_y$. This $B_y$ component can be observed (Figure \ref{fig01}) to be mostly positive (duskward) during the storm main phase and only turns negative (dawnward) towards the latter part of the main phase and stays in this orientation during the recovery phase, but with smaller values. These configurations or shapes are validated statistically by previous studies that used ground-based \citep{sc:65,sc:66}, as well as space-based (e.g, Cluster mission) observations \citep{sc:67,sc:68,sc:69}.

Furthermore, in the terrestrial ionosphere, a distinctive feature under geomagnetically disturbed conditions is the Tongue of Ionization (ToI). It is a tongue-shaped region of enhanced electron density stretching from the mid-latitudes toward the poles. The ToI typically occurs during a geomagnetic storm's main phase. ToIs are often fragmented and form plasma patches over the poles (see \cite{sc:70} and references therein). They are known to trigger strong ionospheric irregularities and subsequent scintillations in communication and satellite-based (GPS) navigation signals over the polar and high latitude regions (see \cite{sc:71} and references therein). ToIs can be probed using TEC, radar, and satellite measurements in addition to the use of the ionospheric tomography technique and physics-driven global circulation model (see \cite{sc:72} and references therein). Although the TEC is closely related to the layer electron density of the $F_2$ layer, its variability is shaped by plasma redistribution through these convection processes. The convection electric field drives large-scale drifts that transport plasma from regions of high production to those of low production. This, in turn, gives rise to features such as the ToI, where enhanced dayside plasma is convected across the polar cap into the nightside. Thus, large-scale TEC patterns do not simply reflect convection, but instead emerge from the combined effects of transport, ionization, and recombination, with the strongest responses occurring when sufficient plasma reservoirs are available (e.g., under sunlit conditions) and diminishing under low-production circumstances such as polar night.

From our observations of the plasma convection due to varying configurations of IMF $B_z$ and $B_y$ during 13 UT on March 23, 2023, we have shown the stations' (palm, mtri, bhrt, and sctb) locations with respect to the two-cell convection pattern over the southern polar region. Now, from Figure \ref{fig01}, it is evident that at 13 UT on March 23, 2023, $B_z$ had been southward while $B_y$ had been duskward. A clockwise rotation of the throat flows had occurred. At the same time, the station mtri, which had shown a drastic TEC enhancement, had been under this intense throat flows region. Over the other stations, which were either away (stations: bhrt and sctb) from the throat flows region or were well outside (station: palm) the auroral oval, such levels of enhancements (as seen over station: mtri) were not observed. However, we do observe decreases in the TEC at this time over bhrt and sctb, which could be attributed to the fact that the local time sectors, over bhrt and sctb, were around evening and post-midnight, respectively. 

Coming to the particle precipitation densities, we can infer that the respective TEC enhancement observed over mtri cannot be due to the result of precipitation of particles, as the station is located far away from peak precipitation levels at 13 UT. Additionally, even when the stations (bhrt and sctb in Figure \ref{fig02} top panel) are under strong particle precipitation or within the auroral oval, such levels of enhancement over these stations are not observed. Therefore, the effects of auroral precipitation on the observed drastic enhancements in the diurnal maximum of TEC can be neglected. Similar trends were observed for the other five events, and based on all six events, a statistical variation in terms of correlation of the convection velocities and the precipitation densities with the TEC enhancements from the respective stations was shown. The correlation coefficients showed that the velocities related to plasma convection were highly correlated with the TEC enhancements in comparison to the particle precipitation densities. These high levels of correlation, during the main phase of the space weather events under consideration, may be attributed to the different magnitudes of the southward and the dawn-dusk IMF, which influence the structures, shapes, orientations, and positions of the plasma convection cells and the corresponding electrodynamical divider.     

It is to be noted that the statistical basis of this study is inherently limited. Because our focus is on the peak-time ionospheric response, only six diurnal maxima from the available six strong/intense storms are analyzed. This event-based approach highlights the geophysical drivers of the largest responses but does not provide the robustness of a large-sample statistical analysis. Therefore, the results should be interpreted as a comparative case-based analysis rather than as statistically rigorous correlations. This limitation does not undermine the qualitative summary that peak TEC enhancements more closely track convection-driven transport than precipitation alone, but it does emphasize the need for caution when generalizing beyond the events studied in this work.

As a path forward for this work, we will investigate the north-south asymmetries in terms of SwMI coupling and the corresponding responses of the northern and southern polar ionospheres simultaneously under a larger database of strong geomagnetic events. These studies would not only become pertinent to physics-driven models in terms of improvements of the same to precisely predict geomagnetic storm-time ionospheric conditions, but also would be pertinent to the development of a reliable space weather forecast system, especially over the less-explored Antarctic region.   

\section{Summary}

The present work was directed at correlating the observed enhancements in the diurnal maximum of TEC with the plasma convection velocity and the particle precipitation densities over the Antarctic region. We considered the main phases of six intense/strong geomagnetic storm-time conditions during the year 2023. Our results suggest that the precipitation densities were less correlated with the observed enhanced TEC variations in comparison to the plasma convection and the corresponding velocities. This may be attributed to the position of the electrodynamical divider and the associated throat flows region, as a result of varying magnitudes and orientations of the north-south and dawn-dusk components of the IMF. Further studies are needed to identify the underlying mechanisms relating ionospheric density perturbations and the physical processes over the under-explored south polar region under varying severity of geomagnetic events.

\section*{Acknowledgments}

S.C. acknowledges the Department of Science and Technology (DST), Government of India, for providing the Research Associate fellowship. This work is part of the project POlar and space WEather Research (POWER) of the Indian Institute of Geomagnetism and is supported by DST, India. The authors acknowledge the NASA OMNIWeb Data Explorer for the openly available high-resolution (1-minute) IMF $B_z$ and $B_y$, the $V_{sw}$, the density, the $IEF_y$, and the SYM-H data. The authors would also like to thank the IIG and NCPOR, MoES staff for their logistical and scientific support in maintaining the database at the two Indian Antarctic stations (bhrt and mtri). Acknowledgments are due to the SuperMag network for providing openly available SML index data. The authors thank the PIs of the magnetic observatories and the national institutes that support the observatories. We thank both reviewers for their valuable suggestions that helped us in enhancing the manuscript quality. The authors thank the IGS network for providing openly available GPS TEC data for our analysis of ionospheric conditions over the Antarctic regions. The NCAR WACCM-X model hosted by the CCMC is duly acknowledged for the simulation runs. The authors acknowledge the use of SuperDARN data. SuperDARN is a collection of radars funded by national scientific funding agencies of Australia, Canada, China, France, Italy, Japan, Norway, South Africa, the United Kingdom, and the United States of America. Acknowledgments go to Patrick Newell of the JHUAPL for developing the Ovation-Prime model hosted at CCMC. 

\bibliographystyle{jasr-model5-names}
\biboptions{authoryear}
\bibliography{SC.bib}

\begin{thebibliography}{67}
\expandafter\ifx\csname natexlab\endcsname\relax\def\natexlab#1{#1}\fi
\ifx\xfnm\relax \def\xfnm[#1]{\unskip,\space#1}\fi

\bibitem[{Akasofu(1981)}]{sc:02}
\bibinfo{author}{Akasofu, S.-I.} (\bibinfo{year}{1981}).
\newblock \bibinfo{title}{{Energy coupling between the solar wind and the magnetosphere}}.
\newblock {\it \bibinfo{journal}{Space Science Reviews}\/},  {\it \bibinfo{volume}{28}\/}\bibinfo{issue}{(2)}, \bibinfo{pages}{121--190}. \URLprefix \url{https://doi.org/10.1007/BF00218810}. \DOIprefix\doi{10.1007/BF00218810}.

\bibitem[{Basu et~al.(2002)Basu, Groves, Basu \& Sultan}]{sc:19}
\bibinfo{author}{Basu, S.}, \bibinfo{author}{Groves, K.}, \bibinfo{author}{Basu, S.} et~al. (\bibinfo{year}{2002}).
\newblock \bibinfo{title}{{Specification and forecasting of scintillations in communication/navigation links: current status and future plans}}.
\newblock {\it \bibinfo{journal}{Journal of Atmospheric and Solar-Terrestrial Physics}\/},  {\it \bibinfo{volume}{64}\/}\bibinfo{issue}{(16)}, \bibinfo{pages}{1745--1754}. \URLprefix \url{https://www.sciencedirect.com/science/article/pii/S1364682602001244}. \DOIprefix\doi{https://doi.org/10.1016/S1364-6826(02)00124-4}.
\newblock \bibinfo{note}{Space Weather Effects on Technological Systems}.

\bibitem[{Cai et~al.(2024)Cai, Aikio, Oyama, Ivchenko, Vanhamäki, Virtanen, Buchert, Mekuriaw \& Zhang}]{sc:b}
\bibinfo{author}{Cai, L.}, \bibinfo{author}{Aikio, A.}, \bibinfo{author}{Oyama, S.} et~al. (\bibinfo{year}{2024}).
\newblock \bibinfo{title}{{Effect of Polar Cap Patches on the High-Latitude Upper Thermospheric Winds}}.
\newblock {\it \bibinfo{journal}{Journal of Geophysical Research: Space Physics}\/},  {\it \bibinfo{volume}{129}\/}\bibinfo{issue}{(8)}, \bibinfo{pages}{e2024JA032819}. \URLprefix \url{https://agupubs.onlinelibrary.wiley.com/doi/abs/10.1029/2024JA032819}. \DOIprefix\doi{https://doi.org/10.1029/2024JA032819}. \href{http://arxiv.org/abs/https://agupubs.onlinelibrary.wiley.com/doi/pdf/10.1029/2024JA032819}{\tt arXiv:https://agupubs.onlinelibrary.wiley.com/doi/pdf/10.1029/2024JA032819}.
\newblock \bibinfo{note}{E2024JA032819 2024JA032819}.

\bibitem[{Carlson(2012)}]{sc:21}
\bibinfo{author}{Carlson, H.~C.} (\bibinfo{year}{2012}).
\newblock \bibinfo{title}{{Sharpening our thinking about polar cap ionospheric patch morphology, research, and mitigation techniques}}.
\newblock {\it \bibinfo{journal}{Radio Science}\/},  {\it \bibinfo{volume}{47}\/}\bibinfo{issue}{(4)}. \URLprefix \url{https://agupubs.onlinelibrary.wiley.com/doi/abs/10.1029/2011RS004946}. \DOIprefix\doi{https://doi.org/10.1029/2011RS004946}. \href{http://arxiv.org/abs/https://agupubs.onlinelibrary.wiley.com/doi/pdf/10.1029/2011RS004946}{\tt arXiv:https://agupubs.onlinelibrary.wiley.com/doi/pdf/10.1029/2011RS004946}.

\bibitem[{Chakraborty \& Chakrabarty(2023)}]{sc:55}
\bibinfo{author}{Chakraborty, S.},  \& \bibinfo{author}{Chakrabarty, D.} (\bibinfo{year}{2023}).
\newblock \bibinfo{title}{{Global Asymmetry in ${\Delta}X$ Variations During the 06 April 2000 Geomagnetic Storm: Relative Roles of IMF $B_z$ and $B_y$}}.
\newblock {\it \bibinfo{journal}{Journal of Geophysical Research: Space Physics}\/},  {\it \bibinfo{volume}{128}\/}\bibinfo{issue}{(2)}, \bibinfo{pages}{e2022JA031047}. \URLprefix \url{https://agupubs.onlinelibrary.wiley.com/doi/abs/10.1029/2022JA031047}. \DOIprefix\doi{https://doi.org/10.1029/2022JA031047}. \href{http://arxiv.org/abs/https://agupubs.onlinelibrary.wiley.com/doi/pdf/10.1029/2022JA031047}{\tt arXiv:https://agupubs.onlinelibrary.wiley.com/doi/pdf/10.1029/2022JA031047}.
\newblock \bibinfo{note}{E2022JA031047 2022JA031047}.

\bibitem[{Chakraborty et~al.(2024)Chakraborty, Chakrabarty, Yadav \& Seemala}]{sc:51}
\bibinfo{author}{Chakraborty, S.}, \bibinfo{author}{Chakrabarty, D.}, \bibinfo{author}{Yadav, A.~K.} et~al. (\bibinfo{year}{2024}).
\newblock \bibinfo{title}{{Influence of ICME-driven magnetic cloud-like and sheath region induced geomagnetic storms in causing anomalous responses of the low-latitude ionosphere: A case study}}.
\newblock {\it \bibinfo{journal}{Advances in Space Research}\/}, . \URLprefix \url{https://www.sciencedirect.com/science/article/pii/S0273117724012912}. \DOIprefix\doi{https://doi.org/10.1016/j.asr.2024.12.045}.

\bibitem[{Chakraborty et~al.(2020)Chakraborty, Ray, Sur, Datta \& Paul}]{sc:49}
\bibinfo{author}{Chakraborty, S.}, \bibinfo{author}{Ray, S.}, \bibinfo{author}{Sur, D.} et~al. (\bibinfo{year}{2020}).
\newblock \bibinfo{title}{{Effects of CME and CIR induced geomagnetic storms on low-latitude ionization over Indian longitudes in terms of neutral dynamics}}.
\newblock {\it \bibinfo{journal}{Advances in Space Research}\/},  {\it \bibinfo{volume}{65}\/}\bibinfo{issue}{(1)}, \bibinfo{pages}{198--213}. \URLprefix \url{https://www.sciencedirect.com/science/article/pii/S0273117719307252}. \DOIprefix\doi{https://doi.org/10.1016/j.asr.2019.09.047}.

\bibitem[{Chisham et~al.(2007)Chisham, Lester, Milan, Freeman, Bristow, Grocott, McWilliams, Ruohoniemi, Yeoman, Dyson, Greenwald, Kikuchi, Pinnock, Rash, Sato, Sofko, Villain \& Walker}]{sc:57}
\bibinfo{author}{Chisham, G.}, \bibinfo{author}{Lester, M.}, \bibinfo{author}{Milan, S.~E.} et~al. (\bibinfo{year}{2007}).
\newblock \bibinfo{title}{{A decade of the Super Dual Auroral Radar Network (SuperDARN): scientific achievements, new techniques and future directions}}.
\newblock {\it \bibinfo{journal}{Surveys in Geophysics}\/},  {\it \bibinfo{volume}{28}\/}\bibinfo{issue}{(1)}, \bibinfo{pages}{33--109}. \URLprefix \url{https://doi.org/10.1007/s10712-007-9017-8}. \DOIprefix\doi{10.1007/s10712-007-9017-8}.

\bibitem[{Cousins et~al.(2015)Cousins, Matsuo \& Richmond}]{sc:66}
\bibinfo{author}{Cousins, E. D.~P.}, \bibinfo{author}{Matsuo, T.},  \& \bibinfo{author}{Richmond, A.~D.} (\bibinfo{year}{2015}).
\newblock \bibinfo{title}{{Mapping high-latitude ionospheric electrodynamics with SuperDARN and AMPERE}}.
\newblock {\it \bibinfo{journal}{Journal of Geophysical Research: Space Physics}\/},  {\it \bibinfo{volume}{120}\/}\bibinfo{issue}{(7)}, \bibinfo{pages}{5854--5870}. \URLprefix \url{https://agupubs.onlinelibrary.wiley.com/doi/abs/10.1002/2014JA020463}. \DOIprefix\doi{https://doi.org/10.1002/2014JA020463}. \href{http://arxiv.org/abs/https://agupubs.onlinelibrary.wiley.com/doi/pdf/10.1002/2014JA020463}{\tt arXiv:https://agupubs.onlinelibrary.wiley.com/doi/pdf/10.1002/2014JA020463}.

\bibitem[{Cousins \& Shepherd(2010)}]{sc:65}
\bibinfo{author}{Cousins, E. D.~P.},  \& \bibinfo{author}{Shepherd, S.~G.} (\bibinfo{year}{2010}).
\newblock \bibinfo{title}{{A dynamical model of high-latitude convection derived from SuperDARN plasma drift measurements}}.
\newblock {\it \bibinfo{journal}{Journal of Geophysical Research: Space Physics}\/},  {\it \bibinfo{volume}{115}\/}\bibinfo{issue}{(A12)}. \URLprefix \url{https://agupubs.onlinelibrary.wiley.com/doi/abs/10.1029/2010JA016017}. \DOIprefix\doi{https://doi.org/10.1029/2010JA016017}. \href{http://arxiv.org/abs/https://agupubs.onlinelibrary.wiley.com/doi/pdf/10.1029/2010JA016017}{\tt arXiv:https://agupubs.onlinelibrary.wiley.com/doi/pdf/10.1029/2010JA016017}.

\bibitem[{Davies(1969)}]{sc:23}
\bibinfo{author}{Davies, K.} (\bibinfo{year}{1969}).
\newblock \bibinfo{title}{{Ionospheric radio waves}}.
\newblock {\it \bibinfo{journal}{Ionospheric radio waves}\/}, .

\bibitem[{Dungey(1961)}]{sc:01}
\bibinfo{author}{Dungey, J.~W.} (\bibinfo{year}{1961}).
\newblock \bibinfo{title}{{Interplanetary Magnetic Field and the Auroral Zones}}.
\newblock {\it \bibinfo{journal}{Phys. Rev. Lett.}\/},  {\it \bibinfo{volume}{6}\/}, \bibinfo{pages}{47--48}. \URLprefix \url{https://link.aps.org/doi/10.1103/PhysRevLett.6.47}. \DOIprefix\doi{10.1103/PhysRevLett.6.47}.

\bibitem[{Estey \& Meertens(1999)}]{sc:61}
\bibinfo{author}{Estey, L.~H.},  \& \bibinfo{author}{Meertens, C.~M.} (\bibinfo{year}{1999}).
\newblock \bibinfo{title}{{TEQC: the multi-purpose toolkit for GPS/GLONASS data}}.
\newblock {\it \bibinfo{journal}{GPS solutions}\/},  {\it \bibinfo{volume}{3}\/}\bibinfo{issue}{(1)}, \bibinfo{pages}{42--49}.

\bibitem[{Fang et~al.(2010)Fang, Randall, Lummerzheim, Wang, Fuller-Rowell, Millan \& Lu}]{sc:i}
\bibinfo{author}{Fang, X.}, \bibinfo{author}{Randall, C.~E.}, \bibinfo{author}{Lummerzheim, D.} et~al. (\bibinfo{year}{2010}).
\newblock \bibinfo{title}{{Electron impact ionization: A new parameterization for 30–100 keV electrons}}.
\newblock {\it \bibinfo{journal}{Journal of Geophysical Research: Space Physics}\/},  {\it \bibinfo{volume}{115}\/}\bibinfo{issue}{(A9)}. \URLprefix \url{https://doi.org/10.1029/2009JA014832}. \DOIprefix\doi{10.1029/2009JA014832}.

\bibitem[{Fedrizzi et~al.(2002)Fedrizzi, de~Paula, Kantor, Langley, Santos \& Komjathy}]{sc:45}
\bibinfo{author}{Fedrizzi, M.}, \bibinfo{author}{de~Paula, E.~R.}, \bibinfo{author}{Kantor, I.~J.} et~al. (\bibinfo{year}{2002}).
\newblock \bibinfo{title}{{Mapping the low-latitude ionosphere with GPS}}.
\newblock {\it \bibinfo{journal}{GPS WORLD}\/},  {\it \bibinfo{volume}{13}\/}\bibinfo{issue}{(2)}, \bibinfo{pages}{41--47}.

\bibitem[{F\"orster et~al.(2007)F\"orster, Paschmann, Haaland, Quinn, Torbert, Vaith \& Kletzing}]{sc:68}
\bibinfo{author}{F\"orster, M.}, \bibinfo{author}{Paschmann, G.}, \bibinfo{author}{Haaland, S.~E.} et~al. (\bibinfo{year}{2007}).
\newblock \bibinfo{title}{{High-latitude plasma convection from Cluster EDI: variances and solar wind correlations}}.
\newblock {\it \bibinfo{journal}{Annales Geophysicae}\/},  {\it \bibinfo{volume}{25}\/}\bibinfo{issue}{(7)}, \bibinfo{pages}{1691--1707}. \URLprefix \url{https://angeo.copernicus.org/articles/25/1691/2007/}. \DOIprefix\doi{10.5194/angeo-25-1691-2007}.

\bibitem[{Foster(1993)}]{sc:30}
\bibinfo{author}{Foster, J.~C.} (\bibinfo{year}{1993}).
\newblock \bibinfo{title}{{Storm time plasma transport at middle and high latitudes}}.
\newblock {\it \bibinfo{journal}{Journal of Geophysical Research: Space Physics}\/},  {\it \bibinfo{volume}{98}\/}\bibinfo{issue}{(A2)}, \bibinfo{pages}{1675--1689}.

\bibitem[{Foster \& Vo(2002)}]{sc:a}
\bibinfo{author}{Foster, J.~C.},  \& \bibinfo{author}{Vo, H.~B.} (\bibinfo{year}{2002}).
\newblock \bibinfo{title}{{Average characteristics and activity dependence of the subauroral polarization stream}}.
\newblock {\it \bibinfo{journal}{Journal of Geophysical Research: Space Physics}\/},  {\it \bibinfo{volume}{107}\/}\bibinfo{issue}{(A12)}, \bibinfo{pages}{SIA 16--1--SIA 16--10}. \URLprefix \url{https://agupubs.onlinelibrary.wiley.com/doi/abs/10.1029/2002JA009409}. \DOIprefix\doi{https://doi.org/10.1029/2002JA009409}. \href{http://arxiv.org/abs/https://agupubs.onlinelibrary.wiley.com/doi/pdf/10.1029/2002JA009409}{\tt arXiv:https://agupubs.onlinelibrary.wiley.com/doi/pdf/10.1029/2002JA009409}.

\bibitem[{Förster \& Cnossen(2013)}]{sc:41}
\bibinfo{author}{Förster, M.},  \& \bibinfo{author}{Cnossen, I.} (\bibinfo{year}{2013}).
\newblock \bibinfo{title}{{Upper atmosphere differences between northern and southern high latitudes: The role of magnetic field asymmetry}}.
\newblock {\it \bibinfo{journal}{Journal of Geophysical Research: Space Physics}\/},  {\it \bibinfo{volume}{118}\/}\bibinfo{issue}{(9)}, \bibinfo{pages}{5951--5966}. \URLprefix \url{https://agupubs.onlinelibrary.wiley.com/doi/abs/10.1002/jgra.50554}. \DOIprefix\doi{https://doi.org/10.1002/jgra.50554}. \href{http://arxiv.org/abs/https://agupubs.onlinelibrary.wiley.com/doi/pdf/10.1002/jgra.50554}{\tt arXiv:https://agupubs.onlinelibrary.wiley.com/doi/pdf/10.1002/jgra.50554}.

\bibitem[{Förster \& Haaland(2015)}]{sc:69}
\bibinfo{author}{Förster, M.},  \& \bibinfo{author}{Haaland, S.} (\bibinfo{year}{2015}).
\newblock \bibinfo{title}{{Interhemispheric differences in ionospheric convection: Cluster EDI observations revisited}}.
\newblock {\it \bibinfo{journal}{Journal of Geophysical Research: Space Physics}\/},  {\it \bibinfo{volume}{120}\/}\bibinfo{issue}{(7)}, \bibinfo{pages}{5805--5823}. \URLprefix \url{https://agupubs.onlinelibrary.wiley.com/doi/abs/10.1002/2014JA020774}. \DOIprefix\doi{https://doi.org/10.1002/2014JA020774}. \href{http://arxiv.org/abs/https://agupubs.onlinelibrary.wiley.com/doi/pdf/10.1002/2014JA020774}{\tt arXiv:https://agupubs.onlinelibrary.wiley.com/doi/pdf/10.1002/2014JA020774}.

\bibitem[{Gjerloev(2009)}]{sc:53}
\bibinfo{author}{Gjerloev, J.~W.} (\bibinfo{year}{2009}).
\newblock \bibinfo{title}{{A Global Ground-Based Magnetometer Initiative}}.
\newblock {\it \bibinfo{journal}{Eos, Transactions American Geophysical Union}\/},  {\it \bibinfo{volume}{90}\/}\bibinfo{issue}{(27)}, \bibinfo{pages}{230--231}. \URLprefix \url{https://agupubs.onlinelibrary.wiley.com/doi/abs/10.1029/2009EO270002}. \DOIprefix\doi{https://doi.org/10.1029/2009EO270002}. \href{http://arxiv.org/abs/https://agupubs.onlinelibrary.wiley.com/doi/pdf/10.1029/2009EO270002}{\tt arXiv:https://agupubs.onlinelibrary.wiley.com/doi/pdf/10.1029/2009EO270002}.

\bibitem[{Gjerloev(2012)}]{sc:54}
\bibinfo{author}{Gjerloev, J.~W.} (\bibinfo{year}{2012}).
\newblock \bibinfo{title}{{The SuperMAG data processing technique}}.
\newblock {\it \bibinfo{journal}{Journal of Geophysical Research: Space Physics}\/},  {\it \bibinfo{volume}{117}\/}\bibinfo{issue}{(A9)}. \URLprefix \url{https://agupubs.onlinelibrary.wiley.com/doi/abs/10.1029/2012JA017683}. \DOIprefix\doi{https://doi.org/10.1029/2012JA017683}. \href{http://arxiv.org/abs/https://agupubs.onlinelibrary.wiley.com/doi/pdf/10.1029/2012JA017683}{\tt arXiv:https://agupubs.onlinelibrary.wiley.com/doi/pdf/10.1029/2012JA017683}.

\bibitem[{Gonzalez et~al.(1994)Gonzalez, Joselyn, Kamide, Kroehl, Rostoker, Tsurutani \& Vasyliunas}]{sc:05}
\bibinfo{author}{Gonzalez, W.~D.}, \bibinfo{author}{Joselyn, J.~A.}, \bibinfo{author}{Kamide, Y.} et~al. (\bibinfo{year}{1994}).
\newblock \bibinfo{title}{{What is a geomagnetic storm?}}
\newblock {\it \bibinfo{journal}{Journal of Geophysical Research: Space Physics}\/},  {\it \bibinfo{volume}{99}\/}\bibinfo{issue}{(A4)}, \bibinfo{pages}{5771--5792}. \URLprefix \url{https://agupubs.onlinelibrary.wiley.com/doi/abs/10.1029/93JA02867}. \DOIprefix\doi{https://doi.org/10.1029/93JA02867}. \href{http://arxiv.org/abs/https://agupubs.onlinelibrary.wiley.com/doi/pdf/10.1029/93JA02867}{\tt arXiv:https://agupubs.onlinelibrary.wiley.com/doi/pdf/10.1029/93JA02867}.

\bibitem[{Gonzalez \& Tsurutani(1987)}]{sc:03}
\bibinfo{author}{Gonzalez, W.~D.},  \& \bibinfo{author}{Tsurutani, B.~T.} (\bibinfo{year}{1987}).
\newblock \bibinfo{title}{{Criteria of interplanetary parameters causing intense magnetic storms (Dst < âˆ’100 nT)}}.
\newblock {\it \bibinfo{journal}{Planetary and Space Science}\/},  {\it \bibinfo{volume}{35}\/}\bibinfo{issue}{(9)}, \bibinfo{pages}{1101--1109}. \URLprefix \url{https://www.sciencedirect.com/science/article/pii/0032063387900158}. \DOIprefix\doi{https://doi.org/10.1016/0032-0633(87)90015-8}.

\bibitem[{Greenwald et~al.(1995)Greenwald, Baker, Dudeney, Pinnock, Jones, Thomas, Villain, Cerisier, Senior, Hanuise, Hunsucker, Sofko, Koehler, Nielsen, Pellinen, Walker, Sato \& Yamagishi}]{sc:56}
\bibinfo{author}{Greenwald, R.~A.}, \bibinfo{author}{Baker, K.~B.}, \bibinfo{author}{Dudeney, J.~R.} et~al. (\bibinfo{year}{1995}).
\newblock \bibinfo{title}{{DARN/SuperDARN}}.
\newblock {\it \bibinfo{journal}{Space Science Reviews}\/},  {\it \bibinfo{volume}{71}\/}\bibinfo{issue}{(1)}, \bibinfo{pages}{761--796}. \URLprefix \url{https://doi.org/10.1007/BF00751350}. \DOIprefix\doi{10.1007/BF00751350}.

\bibitem[{Haaland et~al.(2007)Haaland, Paschmann, F\"orster, Quinn, Torbert, McIlwain, Vaith, Puhl-Quinn \& Kletzing}]{sc:67}
\bibinfo{author}{Haaland, S.~E.}, \bibinfo{author}{Paschmann, G.}, \bibinfo{author}{F\"orster, M.} et~al. (\bibinfo{year}{2007}).
\newblock \bibinfo{title}{{High-latitude plasma convection from Cluster EDI measurements: method and IMF-dependence}}.
\newblock {\it \bibinfo{journal}{Annales Geophysicae}\/},  {\it \bibinfo{volume}{25}\/}\bibinfo{issue}{(1)}, \bibinfo{pages}{239--253}. \URLprefix \url{https://angeo.copernicus.org/articles/25/239/2007/}. \DOIprefix\doi{10.5194/angeo-25-239-2007}.

\bibitem[{Hargreaves(1992)}]{sc:24}
\bibinfo{author}{Hargreaves, J.~K.} (\bibinfo{year}{1992}).
\newblock \bibinfo{title}{{The solar-terrestrial environment: an introduction to geospace-the science of the terrestrial upper atmosphere, ionosphere, and magnetosphere}}, .

\bibitem[{Heelis(1984)}]{sc:l}
\bibinfo{author}{Heelis, R.~A.} (\bibinfo{year}{1984}).
\newblock \bibinfo{title}{{The effects of interplanetary magnetic field orientation on dayside high-latitude convection}}.
\newblock {\it \bibinfo{journal}{Journal of Geophysical Research: Space Physics}\/},  {\it \bibinfo{volume}{89}\/}\bibinfo{issue}{(A5)}, \bibinfo{pages}{2873--2880}. \URLprefix \url{https://doi.org/10.1029/JA089iA05p02873}. \DOIprefix\doi{10.1029/JA089iA05p02873}.

\bibitem[{Huo et~al.(2005)Huo, Yuan, Ou, Wen, Luo et~al.}]{sc:25}
\bibinfo{author}{Huo, X.}, \bibinfo{author}{Yuan, Y.}, \bibinfo{author}{Ou, J.} et~al. (\bibinfo{year}{2005}).
\newblock \bibinfo{title}{{The diurnal variations, semiannual and winter anomalies of the ionospheric TEC based on GPS data in China}}.
\newblock {\it \bibinfo{journal}{Prog. Nat. Sci}\/},  {\it \bibinfo{volume}{15}\/}\bibinfo{issue}{(1)}, \bibinfo{pages}{56--60}.

\bibitem[{Kelley(2009)}]{sc:j}
\bibinfo{author}{Kelley, M.~C.} (\bibinfo{year}{2009}).
\newblock \bibinfo{title}{{The Earth's Ionosphere: Plasma Physics and Electrodynamics}}, .
\newblock \URLprefix \url{https://www.elsevier.com/books/the-earths-ionosphere/kelley/978-0-12-088425-4}.

\bibitem[{Kintner et~al.(2007)Kintner, Ledvina \& de~Paula}]{sc:20}
\bibinfo{author}{Kintner, P.~M.}, \bibinfo{author}{Ledvina, B.~M.},  \& \bibinfo{author}{de~Paula, E.~R.} (\bibinfo{year}{2007}).
\newblock \bibinfo{title}{{GPS and ionospheric scintillations}}.
\newblock {\it \bibinfo{journal}{Space Weather}\/},  {\it \bibinfo{volume}{5}\/}\bibinfo{issue}{(9)}. \URLprefix \url{https://agupubs.onlinelibrary.wiley.com/doi/abs/10.1029/2006SW000260}. \DOIprefix\doi{https://doi.org/10.1029/2006SW000260}. \href{http://arxiv.org/abs/https://agupubs.onlinelibrary.wiley.com/doi/pdf/10.1029/2006SW000260}{\tt arXiv:https://agupubs.onlinelibrary.wiley.com/doi/pdf/10.1029/2006SW000260}.

\bibitem[{Laundal et~al.(2017)Laundal, Cnossen, Milan, Haaland, Coxon, Pedatella, F{\"o}rster \& Reistad}]{sc:42}
\bibinfo{author}{Laundal, K.~M.}, \bibinfo{author}{Cnossen, I.}, \bibinfo{author}{Milan, S.~E.} et~al. (\bibinfo{year}{2017}).
\newblock \bibinfo{title}{{North--South Asymmetries in Earth's Magnetic Field}}.
\newblock {\it \bibinfo{journal}{Space Science Reviews}\/},  {\it \bibinfo{volume}{206}\/}\bibinfo{issue}{(1)}, \bibinfo{pages}{225--257}. \URLprefix \url{https://doi.org/10.1007/s11214-016-0273-0}. \DOIprefix\doi{10.1007/s11214-016-0273-0}.

\bibitem[{Liu et~al.(2018)Liu, Bardeen, Foster, Lauritzen, Liu, Lu, Marsh, Maute, McInerney, Pedatella, Qian, Richmond, Roble, Solomon, Vitt \& Wang}]{sc:48}
\bibinfo{author}{Liu, H.-L.}, \bibinfo{author}{Bardeen, C.~G.}, \bibinfo{author}{Foster, B.~T.} et~al. (\bibinfo{year}{2018}).
\newblock \bibinfo{title}{{Development and Validation of the Whole Atmosphere Community Climate Model With Thermosphere and Ionosphere Extension (WACCM-X 2.0)}}.
\newblock {\it \bibinfo{journal}{Journal of Advances in Modeling Earth Systems}\/},  {\it \bibinfo{volume}{10}\/}\bibinfo{issue}{(2)}, \bibinfo{pages}{381--402}. \URLprefix \url{https://agupubs.onlinelibrary.wiley.com/doi/abs/10.1002/2017MS001232}. \DOIprefix\doi{https://doi.org/10.1002/2017MS001232}. \href{http://arxiv.org/abs/https://agupubs.onlinelibrary.wiley.com/doi/pdf/10.1002/2017MS001232}{\tt arXiv:https://agupubs.onlinelibrary.wiley.com/doi/pdf/10.1002/2017MS001232}.

\bibitem[{Liu et~al.(2010)Liu, Foster, Hagan, McInerney, Maute, Qian, Richmond, Roble, Solomon, Garcia, Kinnison, Marsh, Smith, Richter, Sassi \& Oberheide}]{sc:47}
\bibinfo{author}{Liu, H.-L.}, \bibinfo{author}{Foster, B.~T.}, \bibinfo{author}{Hagan, M.~E.} et~al. (\bibinfo{year}{2010}).
\newblock \bibinfo{title}{{Thermosphere extension of the Whole Atmosphere Community Climate Model}}.
\newblock {\it \bibinfo{journal}{Journal of Geophysical Research: Space Physics}\/},  {\it \bibinfo{volume}{115}\/}\bibinfo{issue}{(A12)}. \URLprefix \url{https://agupubs.onlinelibrary.wiley.com/doi/abs/10.1029/2010JA015586}. \DOIprefix\doi{https://doi.org/10.1029/2010JA015586}. \href{http://arxiv.org/abs/https://agupubs.onlinelibrary.wiley.com/doi/pdf/10.1029/2010JA015586}{\tt arXiv:https://agupubs.onlinelibrary.wiley.com/doi/pdf/10.1029/2010JA015586}.

\bibitem[{Liu et~al.(2009)Liu, Wan, Ning \& Zhang}]{sc:28}
\bibinfo{author}{Liu, L.}, \bibinfo{author}{Wan, W.}, \bibinfo{author}{Ning, B.} et~al. (\bibinfo{year}{2009}).
\newblock \bibinfo{title}{{Climatology of the mean total electron content derived from GPS global ionospheric maps}}.
\newblock {\it \bibinfo{journal}{Journal of Geophysical Research: Space Physics}\/},  {\it \bibinfo{volume}{114}\/}\bibinfo{issue}{(A6)}. \URLprefix \url{https://agupubs.onlinelibrary.wiley.com/doi/abs/10.1029/2009JA014244}. \DOIprefix\doi{https://doi.org/10.1029/2009JA014244}. \href{http://arxiv.org/abs/https://agupubs.onlinelibrary.wiley.com/doi/pdf/10.1029/2009JA014244}{\tt arXiv:https://agupubs.onlinelibrary.wiley.com/doi/pdf/10.1029/2009JA014244}.

\bibitem[{{Loewe, C. A. and Prölss, G. W.}(1997)}]{sc:06}
\bibinfo{author}{{Loewe, C. A. and Prölss, G. W.}} (\bibinfo{year}{1997}).
\newblock \bibinfo{title}{{Classification and mean behavior of magnetic storms}}.
\newblock {\it \bibinfo{journal}{Journal of Geophysical Research: Space Physics}\/},  {\it \bibinfo{volume}{102}\/}\bibinfo{issue}{(A7)}, \bibinfo{pages}{14209--14213}. \URLprefix \url{https://agupubs.onlinelibrary.wiley.com/doi/abs/10.1029/96JA04020}. \DOIprefix\doi{https://doi.org/10.1029/96JA04020}. \href{http://arxiv.org/abs/https://agupubs.onlinelibrary.wiley.com/doi/pdf/10.1029/96JA04020}{\tt arXiv:https://agupubs.onlinelibrary.wiley.com/doi/pdf/10.1029/96JA04020}.

\bibitem[{Ma \& Schunk(1997)}]{sc:c}
\bibinfo{author}{Ma, T.-Z.},  \& \bibinfo{author}{Schunk, R.} (\bibinfo{year}{1997}).
\newblock \bibinfo{title}{{Effect of polar cap patches on the thermosphere for different solar activity levels}}.
\newblock {\it \bibinfo{journal}{Journal of Atmospheric and Solar-Terrestrial Physics}\/},  {\it \bibinfo{volume}{59}\/}\bibinfo{issue}{(14)}, \bibinfo{pages}{1823--1829}. \URLprefix \url{https://www.sciencedirect.com/science/article/pii/S1364682697000035}. \DOIprefix\doi{https://doi.org/10.1016/S1364-6826(97)00003-5}.

\bibitem[{Mannucci et~al.(1993)Mannucci, Wilson \& Edwards}]{sc:44}
\bibinfo{author}{Mannucci, A.~J.}, \bibinfo{author}{Wilson, B.~D.},  \& \bibinfo{author}{Edwards, C.~D.} (\bibinfo{year}{1993}).
\newblock \bibinfo{title}{{A new method for monitoring the Earth's ionospheric total electron content using the GPS global network}}, .
\newblock (pp. \bibinfo{pages}{1323--1332}).

\bibitem[{Mendillo(2006)}]{sc:26}
\bibinfo{author}{Mendillo, M.} (\bibinfo{year}{2006}).
\newblock \bibinfo{title}{{Storms in the ionosphere: Patterns and processes for total electron content}}.
\newblock {\it \bibinfo{journal}{Reviews of Geophysics}\/},  {\it \bibinfo{volume}{44}\/}\bibinfo{issue}{(4)}.

\bibitem[{Moen et~al.(2013)Moen, Oksavik, Alfonsi, Daabakk, Romano \& Spogli}]{sc:22}
\bibinfo{author}{Moen, J.}, \bibinfo{author}{Oksavik, K.}, \bibinfo{author}{Alfonsi, L.} et~al. (\bibinfo{year}{2013}).
\newblock \bibinfo{title}{{Space weather challenges of the polar cap ionosphere}}.
\newblock {\it \bibinfo{journal}{J. Space Weather Space Clim.}\/},  {\it \bibinfo{volume}{3}\/}, \bibinfo{pages}{A02}. \URLprefix \url{https://doi.org/10.1051/swsc/2013025}. \DOIprefix\doi{10.1051/swsc/2013025}.

\bibitem[{Newell et~al.(2009{\natexlab{a}})Newell, Liou \& Wilson}]{sc:16}
\bibinfo{author}{Newell, P.~T.}, \bibinfo{author}{Liou, K.},  \& \bibinfo{author}{Wilson, G.~R.} (\bibinfo{year}{2009}{\natexlab{a}}).
\newblock \bibinfo{title}{Polar cap particle precipitation and aurora: Review and commentary}.
\newblock {\it \bibinfo{journal}{Journal of Atmospheric and Solar-Terrestrial Physics}\/},  {\it \bibinfo{volume}{71}\/}\bibinfo{issue}{(2)}, \bibinfo{pages}{199--215}. \URLprefix \url{https://www.sciencedirect.com/science/article/pii/S1364682608003738}. \DOIprefix\doi{https://doi.org/10.1016/j.jastp.2008.11.004}.

\bibitem[{Newell et~al.(2009{\natexlab{b}})Newell, Sotirelis \& Wing}]{sc:59}
\bibinfo{author}{Newell, P.~T.}, \bibinfo{author}{Sotirelis, T.},  \& \bibinfo{author}{Wing, S.} (\bibinfo{year}{2009}{\natexlab{b}}).
\newblock \bibinfo{title}{{Diffuse, monoenergetic, and broadband aurora: The global precipitation budget}}.
\newblock {\it \bibinfo{journal}{Journal of Geophysical Research: Space Physics}\/},  {\it \bibinfo{volume}{114}\/}\bibinfo{issue}{(A9)}. \URLprefix \url{https://agupubs.onlinelibrary.wiley.com/doi/abs/10.1029/2009JA014326}. \DOIprefix\doi{https://doi.org/10.1029/2009JA014326}. \href{http://arxiv.org/abs/https://agupubs.onlinelibrary.wiley.com/doi/pdf/10.1029/2009JA014326}{\tt arXiv:https://agupubs.onlinelibrary.wiley.com/doi/pdf/10.1029/2009JA014326}.

\bibitem[{Newell et~al.(2010)Newell, Sotirelis \& Wing}]{sc:60}
\bibinfo{author}{Newell, P.~T.}, \bibinfo{author}{Sotirelis, T.},  \& \bibinfo{author}{Wing, S.} (\bibinfo{year}{2010}).
\newblock \bibinfo{title}{{Seasonal variations in diffuse, monoenergetic, and broadband aurora}}.
\newblock {\it \bibinfo{journal}{Journal of Geophysical Research: Space Physics}\/},  {\it \bibinfo{volume}{115}\/}\bibinfo{issue}{(A3)}. \URLprefix \url{https://agupubs.onlinelibrary.wiley.com/doi/abs/10.1029/2009JA014805}. \DOIprefix\doi{https://doi.org/10.1029/2009JA014805}. \href{http://arxiv.org/abs/https://agupubs.onlinelibrary.wiley.com/doi/pdf/10.1029/2009JA014805}{\tt arXiv:https://agupubs.onlinelibrary.wiley.com/doi/pdf/10.1029/2009JA014805}.

\bibitem[{Nishitani et~al.(2019)Nishitani, Ruohoniemi, Lester, Baker, Koustov, Shepherd, Chisham, Hori, Thomas, Makarevich, Marchaudon, Ponomarenko, Wild, Milan, Bristow, Devlin, Miller, Greenwald, Ogawa \& Kikuchi}]{sc:58}
\bibinfo{author}{Nishitani, N.}, \bibinfo{author}{Ruohoniemi, J.~M.}, \bibinfo{author}{Lester, M.} et~al. (\bibinfo{year}{2019}).
\newblock \bibinfo{title}{{Review of the accomplishments of mid-latitude Super Dual Auroral Radar Network (SuperDARN) HF radars}}.
\newblock {\it \bibinfo{journal}{Progress in Earth and Planetary Science}\/},  {\it \bibinfo{volume}{6}\/}\bibinfo{issue}{(1)}, \bibinfo{pages}{27}. \URLprefix \url{https://doi.org/10.1186/s40645-019-0270-5}. \DOIprefix\doi{10.1186/s40645-019-0270-5}.

\bibitem[{Pedersen et~al.(2024)Pedersen, Juusola, Vanhamäki, Aikio \& Viljanen}]{sc:50}
\bibinfo{author}{Pedersen, M.~N.}, \bibinfo{author}{Juusola, L.}, \bibinfo{author}{Vanhamäki, H.} et~al. (\bibinfo{year}{2024}).
\newblock \bibinfo{title}{{Rapid Geomagnetic Variations During High-Speed Stream, Sheath and Magnetic Cloud-Driven Geomagnetic Storms From 1996 to 2023}}.
\newblock {\it \bibinfo{journal}{Journal of Geophysical Research: Space Physics}\/},  {\it \bibinfo{volume}{129}\/}\bibinfo{issue}{(10)}, \bibinfo{pages}{e2024JA032656}. \URLprefix \url{https://agupubs.onlinelibrary.wiley.com/doi/abs/10.1029/2024JA032656}. \DOIprefix\doi{https://doi.org/10.1029/2024JA032656}. \href{http://arxiv.org/abs/https://agupubs.onlinelibrary.wiley.com/doi/pdf/10.1029/2024JA032656}{\tt arXiv:https://agupubs.onlinelibrary.wiley.com/doi/pdf/10.1029/2024JA032656}.
\newblock \bibinfo{note}{E2024JA032656 2024JA032656}.

\bibitem[{Pokhotelov et~al.(2021)Pokhotelov, Fernandez-Gomez \& Borries}]{sc:71}
\bibinfo{author}{Pokhotelov, D.}, \bibinfo{author}{Fernandez-Gomez, I.},  \& \bibinfo{author}{Borries, C.} (\bibinfo{year}{2021}).
\newblock \bibinfo{title}{{Polar tongue of ionisation during geomagnetic superstorm}}.
\newblock {\it \bibinfo{journal}{Annales Geophysicae}\/},  {\it \bibinfo{volume}{39}\/}\bibinfo{issue}{(5)}, \bibinfo{pages}{833--847}. \URLprefix \url{https://angeo.copernicus.org/articles/39/833/2021/}. \DOIprefix\doi{10.5194/angeo-39-833-2021}.

\bibitem[{Rama~Rao et~al.(2006)Rama~Rao, Niranjan, Prasad, Gopi~Krishna \& Uma}]{sc:46}
\bibinfo{author}{Rama~Rao, P. V.~S.}, \bibinfo{author}{Niranjan, K.}, \bibinfo{author}{Prasad, D. S. V. V.~D.} et~al. (\bibinfo{year}{2006}).
\newblock \bibinfo{title}{{On the validity of the ionospheric pierce point (IPP) altitude of 350 km in the Indian equatorial and low-latitude sector}}.
\newblock {\it \bibinfo{journal}{Annales Geophysicae}\/},  {\it \bibinfo{volume}{24}\/}\bibinfo{issue}{(8)}, \bibinfo{pages}{2159--2168}. \URLprefix \url{https://angeo.copernicus.org/articles/24/2159/2006/}. \DOIprefix\doi{10.5194/angeo-24-2159-2006}.

\bibitem[{Roble \& Ridley(1987)}]{sc:h}
\bibinfo{author}{Roble, R.~G.},  \& \bibinfo{author}{Ridley, E.~C.} (\bibinfo{year}{1987}).
\newblock \bibinfo{title}{{An auroral model for the NCAR thermospheric general circulation model (TGCM)}}.
\newblock {\it \bibinfo{journal}{Annales Geophysicae}\/},  {\it \bibinfo{volume}{5A}\/}, \bibinfo{pages}{369--382}. \URLprefix \url{https://www.ann-geophys.net/5/369/1987/}.

\bibitem[{Rodríguez-Zuluaga et~al.(2016)Rodríguez-Zuluaga, Radicella, Nava, Amory-Mazaudier, Mora-Páez \& Alazo-Cuartas}]{sc:52}
\bibinfo{author}{Rodríguez-Zuluaga, J.}, \bibinfo{author}{Radicella, S.~M.}, \bibinfo{author}{Nava, B.} et~al. (\bibinfo{year}{2016}).
\newblock \bibinfo{title}{{Distinct responses of the low-latitude ionosphere to CME and HSSWS: The role of the IMF B oscillation frequency}}.
\newblock {\it \bibinfo{journal}{Journal of Geophysical Research: Space Physics}\/},  {\it \bibinfo{volume}{121}\/}\bibinfo{issue}{(11)}, \bibinfo{pages}{11,528--11,548}. \URLprefix \url{https://agupubs.onlinelibrary.wiley.com/doi/abs/10.1002/2016JA022539}. \DOIprefix\doi{https://doi.org/10.1002/2016JA022539}. \href{http://arxiv.org/abs/https://agupubs.onlinelibrary.wiley.com/doi/pdf/10.1002/2016JA022539}{\tt arXiv:https://agupubs.onlinelibrary.wiley.com/doi/pdf/10.1002/2016JA022539}.

\bibitem[{Ruohoniemi \& Baker(1998)}]{sc:31}
\bibinfo{author}{Ruohoniemi, J.~M.},  \& \bibinfo{author}{Baker, K.~B.} (\bibinfo{year}{1998}).
\newblock \bibinfo{title}{{Large-scale imaging of high-latitude convection with Super Dual Auroral Radar Network HF radar observations}}.
\newblock {\it \bibinfo{journal}{Journal of Geophysical Research: Space Physics}\/},  {\it \bibinfo{volume}{103}\/}\bibinfo{issue}{(A9)}, \bibinfo{pages}{20797--20811}. \URLprefix \url{https://agupubs.onlinelibrary.wiley.com/doi/abs/10.1029/98JA01288}. \DOIprefix\doi{https://doi.org/10.1029/98JA01288}. \href{http://arxiv.org/abs/https://agupubs.onlinelibrary.wiley.com/doi/pdf/10.1029/98JA01288}{\tt arXiv:https://agupubs.onlinelibrary.wiley.com/doi/pdf/10.1029/98JA01288}.

\bibitem[{Schunk \& Nagy(2009)}]{sc:k}
\bibinfo{author}{Schunk, R.~W.},  \& \bibinfo{author}{Nagy, A.~F.} (\bibinfo{year}{2009}).
\newblock \bibinfo{title}{{Ionospheres: Physics, Plasma Physics, and Chemistry}}, .
\newblock \URLprefix \url{https://doi.org/10.1017/CBO9780511635342}.

\bibitem[{Seemala(2023)}]{sc:64}
\bibinfo{author}{Seemala, G.~K.} (\bibinfo{year}{2023}).
\newblock \bibinfo{title}{{Chapter 4 - Estimation of ionospheric total electron content (TEC) from GNSS observations}, editor = {Abhay {Kumar Singh} and Shani Tiwari}, booktitle = {Atmospheric Remote Sensing}}, .
\newblock (pp. \bibinfo{pages}{63--84}). \URLprefix \url{https://www.sciencedirect.com/science/article/pii/B9780323992626000225}. \DOIprefix\doi{https://doi.org/10.1016/B978-0-323-99262-6.00022-5}.

\bibitem[{Seemala et~al.(2023)Seemala, Katual, Kapil \& Vichare}]{sc:29}
\bibinfo{author}{Seemala, G.~K.}, \bibinfo{author}{Katual, I.}, \bibinfo{author}{Kapil, C.} et~al. (\bibinfo{year}{2023}).
\newblock \bibinfo{title}{{Seasonal and solar activity dependence of TEC over Bharati station, Antarctica}}.
\newblock {\it \bibinfo{journal}{Polar Science}\/},  {\it \bibinfo{volume}{38}\/}, \bibinfo{pages}{101001}. \URLprefix \url{https://www.sciencedirect.com/science/article/pii/S1873965223001068}. \DOIprefix\doi{https://doi.org/10.1016/j.polar.2023.101001}.
\newblock \bibinfo{note}{Research Advances from Larsemann Hills, Antarctica: International Cooperation and Future Prospects - Part 1}.

\bibitem[{Seemala \& Valladares(2011)}]{sc:43}
\bibinfo{author}{Seemala, G.~K.},  \& \bibinfo{author}{Valladares, C.~E.} (\bibinfo{year}{2011}).
\newblock \bibinfo{title}{{Statistics of total electron content depletions observed over the South American continent for the year 2008}}.
\newblock {\it \bibinfo{journal}{Radio Science}\/},  {\it \bibinfo{volume}{46}\/}\bibinfo{issue}{(5)}. \URLprefix \url{https://agupubs.onlinelibrary.wiley.com/doi/abs/10.1029/2011RS004722}. \DOIprefix\doi{https://doi.org/10.1029/2011RS004722}. \href{http://arxiv.org/abs/https://agupubs.onlinelibrary.wiley.com/doi/pdf/10.1029/2011RS004722}{\tt arXiv:https://agupubs.onlinelibrary.wiley.com/doi/pdf/10.1029/2011RS004722}.

\bibitem[{Shagimuratov et~al.(2012)Shagimuratov, Krankowski, Ephishov, Cherniak, Wielgosz \& Zakharenkova}]{sc:f}
\bibinfo{author}{Shagimuratov, I.~I.}, \bibinfo{author}{Krankowski, A.}, \bibinfo{author}{Ephishov, I.} et~al. (\bibinfo{year}{2012}).
\newblock \bibinfo{title}{{High latitude TEC fluctuations and irregularity oval during geomagnetic storms}}.
\newblock {\it \bibinfo{journal}{Earth, Planets and Space}\/},  {\it \bibinfo{volume}{64}\/}\bibinfo{issue}{(6)}, \bibinfo{pages}{521--529}. \URLprefix \url{https://doi.org/10.5047/eps.2011.10.015}. \DOIprefix\doi{10.5047/eps.2011.10.015}.

\bibitem[{Shan et~al.(2022)Shan, Yao, Kong, Zhai, Zhou \& Chen}]{sc:72}
\bibinfo{author}{Shan, L.~L.}, \bibinfo{author}{Yao, Y.~B.}, \bibinfo{author}{Kong, J.} et~al. (\bibinfo{year}{2022}).
\newblock \bibinfo{title}{{Three-Dimensional Reconstruction of Tongue of Ionization During the 11 October 2010 Geomagnetic Storm and Evolution Analysis With TIEGCM}}.
\newblock {\it \bibinfo{journal}{Space Weather}\/},  {\it \bibinfo{volume}{20}\/}\bibinfo{issue}{(4)}, \bibinfo{pages}{e2021SW002862}. \URLprefix \url{https://agupubs.onlinelibrary.wiley.com/doi/abs/10.1029/2021SW002862}. \DOIprefix\doi{https://doi.org/10.1029/2021SW002862}. \href{http://arxiv.org/abs/https://agupubs.onlinelibrary.wiley.com/doi/pdf/10.1029/2021SW002862}{\tt arXiv:https://agupubs.onlinelibrary.wiley.com/doi/pdf/10.1029/2021SW002862}.
\newblock \bibinfo{note}{E2021SW002862 2021SW002862}.

\bibitem[{Shreedevi et~al.(2019)Shreedevi, Choudhary, Yu \& Thomas}]{sc:g}
\bibinfo{author}{Shreedevi, P.}, \bibinfo{author}{Choudhary, R.}, \bibinfo{author}{Yu, Y.} et~al. (\bibinfo{year}{2019}).
\newblock \bibinfo{title}{{Morphological study on the ionospheric variability at Bharati, a polar cusp station in the southern hemisphere}}.
\newblock {\it \bibinfo{journal}{Journal of Atmospheric and Solar-Terrestrial Physics}\/},  {\it \bibinfo{volume}{193}\/}, \bibinfo{pages}{105058}. \URLprefix \url{https://www.sciencedirect.com/science/article/pii/S1364682618305996}. \DOIprefix\doi{https://doi.org/10.1016/j.jastp.2019.105058}.

\bibitem[{Thomas \& Shepherd(2018)}]{sc:62}
\bibinfo{author}{Thomas, E.~G.},  \& \bibinfo{author}{Shepherd, S.~G.} (\bibinfo{year}{2018}).
\newblock \bibinfo{title}{{Statistical Patterns of Ionospheric Convection Derived From Mid-latitude, High-Latitude, and Polar SuperDARN HF Radar Observations}}.
\newblock {\it \bibinfo{journal}{Journal of Geophysical Research: Space Physics}\/},  {\it \bibinfo{volume}{123}\/}\bibinfo{issue}{(4)}, \bibinfo{pages}{3196--3216}. \URLprefix \url{https://agupubs.onlinelibrary.wiley.com/doi/abs/10.1002/2018JA025280}. \DOIprefix\doi{https://doi.org/10.1002/2018JA025280}. \href{http://arxiv.org/abs/https://agupubs.onlinelibrary.wiley.com/doi/pdf/10.1002/2018JA025280}{\tt arXiv:https://agupubs.onlinelibrary.wiley.com/doi/pdf/10.1002/2018JA025280}.

\bibitem[{Tsunoda(1988)}]{sc:18}
\bibinfo{author}{Tsunoda, R.~T.} (\bibinfo{year}{1988}).
\newblock \bibinfo{title}{{High-latitude F region irregularities: A review and synthesis}}.
\newblock {\it \bibinfo{journal}{Reviews of Geophysics}\/},  {\it \bibinfo{volume}{26}\/}\bibinfo{issue}{(4)}, \bibinfo{pages}{719--760}.

\bibitem[{Tsurutani et~al.(1992)Tsurutani, Gonzalez, Tang \& Lee}]{sc:04}
\bibinfo{author}{Tsurutani, B.~T.}, \bibinfo{author}{Gonzalez, W.~D.}, \bibinfo{author}{Tang, F.} et~al. (\bibinfo{year}{1992}).
\newblock \bibinfo{title}{{Great magnetic storms}}.
\newblock {\it \bibinfo{journal}{Geophysical Research Letters}\/},  {\it \bibinfo{volume}{19}\/}\bibinfo{issue}{(1)}, \bibinfo{pages}{73--76}. \URLprefix \url{https://agupubs.onlinelibrary.wiley.com/doi/abs/10.1029/91GL02783}. \DOIprefix\doi{https://doi.org/10.1029/91GL02783}. \href{http://arxiv.org/abs/https://agupubs.onlinelibrary.wiley.com/doi/pdf/10.1029/91GL02783}{\tt arXiv:https://agupubs.onlinelibrary.wiley.com/doi/pdf/10.1029/91GL02783}.

\bibitem[{Watson et~al.(2016)Watson, Jayachandran \& MacDougall}]{sc:32}
\bibinfo{author}{Watson, C.}, \bibinfo{author}{Jayachandran, P.~T.},  \& \bibinfo{author}{MacDougall, J.~W.} (\bibinfo{year}{2016}).
\newblock \bibinfo{title}{{Characteristics of GPS TEC variations in the polar cap ionosphere}}.
\newblock {\it \bibinfo{journal}{Journal of Geophysical Research: Space Physics}\/},  {\it \bibinfo{volume}{121}\/}\bibinfo{issue}{(5)}, \bibinfo{pages}{4748--4768}. \URLprefix \url{https://agupubs.onlinelibrary.wiley.com/doi/abs/10.1002/2015JA022275}. \DOIprefix\doi{https://doi.org/10.1002/2015JA022275}. \href{http://arxiv.org/abs/https://agupubs.onlinelibrary.wiley.com/doi/pdf/10.1002/2015JA022275}{\tt arXiv:https://agupubs.onlinelibrary.wiley.com/doi/pdf/10.1002/2015JA022275}.

\bibitem[{Weimer(2005)}]{sc:m}
\bibinfo{author}{Weimer, D.~R.} (\bibinfo{year}{2005}).
\newblock \bibinfo{title}{{Improved ionospheric electrodynamic models and application to calculating Joule heating rates}}.
\newblock {\it \bibinfo{journal}{Journal of Geophysical Research: Space Physics}\/},  {\it \bibinfo{volume}{110}\/}, \bibinfo{pages}{A05306}. \URLprefix \url{https://doi.org/10.1029/2004JA010884}. \DOIprefix\doi{10.1029/2004JA010884}.

\bibitem[{Woods et~al.(1979)Woods, Scourfield, Boynton \& Roach}]{sc:d}
\bibinfo{author}{Woods, A.}, \bibinfo{author}{Scourfield, M.}, \bibinfo{author}{Boynton, D.} et~al. (\bibinfo{year}{1979}).
\newblock \bibinfo{title}{{Plasmasphere convection patterns observed simultaneously from two ground stations}}.
\newblock {\it \bibinfo{journal}{Planetary and Space Science}\/},  {\it \bibinfo{volume}{27}\/}\bibinfo{issue}{(5)}, \bibinfo{pages}{643--652}. \URLprefix \url{https://www.sciencedirect.com/science/article/pii/0032063379901612}. \DOIprefix\doi{https://doi.org/10.1016/0032-0633(79)90161-2}.

\bibitem[{Wu et~al.(2006)Wu, Zhang, Yuan, Wu et~al.}]{sc:27}
\bibinfo{author}{Wu, S.}, \bibinfo{author}{Zhang, K.}, \bibinfo{author}{Yuan, Y.} et~al. (\bibinfo{year}{2006}).
\newblock \bibinfo{title}{{Spatio-temporal characteristics of the ionospheric TEC variation for GPSnet-based real-time positioning in Victoria}}.
\newblock {\it \bibinfo{journal}{Positioning}\/},  {\it \bibinfo{volume}{1}\/}\bibinfo{issue}{(10)}.

\bibitem[{Yahnin \& Sergeev(1996)}]{sc:e}
\bibinfo{author}{Yahnin, A.},  \& \bibinfo{author}{Sergeev, V.} (\bibinfo{year}{1996}).
\newblock \bibinfo{title}{{Simultaneous satellite and ground-based observations of polar cap aurora}}.
\newblock {\it \bibinfo{journal}{Advances in Space Research}\/},  {\it \bibinfo{volume}{18}\/}\bibinfo{issue}{(8)}, \bibinfo{pages}{111--114}. \URLprefix \url{https://www.sciencedirect.com/science/article/pii/0273117795009779}. \DOIprefix\doi{https://doi.org/10.1016/0273-1177(95)00977-9}.
\newblock \bibinfo{note}{The Three-dimensional Magnetosphere}.

\bibitem[{Yeh \& Liu(1982)}]{sc:17}
\bibinfo{author}{Yeh, K.~C.},  \& \bibinfo{author}{Liu, C.-H.} (\bibinfo{year}{1982}).
\newblock \bibinfo{title}{{Radio wave scintillations in the ionosphere}}.
\newblock {\it \bibinfo{journal}{Proceedings of the IEEE}\/},  {\it \bibinfo{volume}{70}\/}\bibinfo{issue}{(4)}, \bibinfo{pages}{324--360}.

\bibitem[{Zhang et~al.(2021)Zhang, Wang, Liu, Zheng, He, Gao, Sun \& Zhong}]{sc:70}
\bibinfo{author}{Zhang, K.}, \bibinfo{author}{Wang, H.}, \bibinfo{author}{Liu, J.} et~al. (\bibinfo{year}{2021}).
\newblock \bibinfo{title}{{Dynamics of the Tongue of Ionizations During the Geomagnetic Storm on September 7, 2015}}.
\newblock {\it \bibinfo{journal}{Journal of Geophysical Research: Space Physics}\/},  {\it \bibinfo{volume}{126}\/}\bibinfo{issue}{(6)}, \bibinfo{pages}{e2020JA029038}. \URLprefix \url{https://agupubs.onlinelibrary.wiley.com/doi/abs/10.1029/2020JA029038}. \DOIprefix\doi{https://doi.org/10.1029/2020JA029038}. \href{http://arxiv.org/abs/https://agupubs.onlinelibrary.wiley.com/doi/pdf/10.1029/2020JA029038}{\tt arXiv:https://agupubs.onlinelibrary.wiley.com/doi/pdf/10.1029/2020JA029038}.
\newblock \bibinfo{note}{E2020JA029038 2020JA029038}.

\end{thebibliography}

\end{document}